\documentclass[aps,prb,twocolumn,amsmath,amssymb,nofootinbib,superscriptaddress,floatfix,reprint,longbibliography]{revtex4-1}
\usepackage[dvips]{graphicx}
\usepackage{latexsym}
\usepackage{amsmath}
\usepackage{amsfonts}
\usepackage{amssymb}
\usepackage{bm}
\usepackage{color}
\usepackage{txfonts}
\usepackage{float}
\usepackage{url}
\usepackage[colorlinks=true, urlcolor=blue, linkcolor=blue, citecolor=blue, pdftex]{hyperref}
\usepackage{ulem}
\usepackage{physics}
\usepackage{hhline}
\begin{document}
	\newcommand{\fig}[2]{\includegraphics[width=#1]{#2}}
	\newcommand{\pprl}{Phys. Rev. Lett. \ }
	\newcommand{\pprb}{Phys. Rev. {B}}

\title {Interband pairing as the origin of the sublattice dichotomy in monolayer FeSe/SrTiO$_3$}

\author{Zhipeng Xu}
\affiliation{Beijing National Laboratory for Condensed Matter Physics and Institute of Physics,
	Chinese Academy of Sciences, Beijing 100190, China}
\affiliation{School of Physical Sciences, University of Chinese Academy of Sciences, Beijing 100190, China}

\author{Shengshan Qin}\email{qinshengshan@bit.edu.cn}
\affiliation{School of Physics, Beijing Institute of Technology, Beijing 100081, China}

\author{Kun Jiang}
\email{jiangkun@iphy.ac.cn}
\affiliation{Beijing National Laboratory for Condensed Matter Physics and Institute of Physics,
	Chinese Academy of Sciences, Beijing 100190, China}
\affiliation{School of Physical Sciences, University of Chinese Academy of Sciences, Beijing 100190, China}

\author{Jiangping Hu}
\email{jphu@iphy.ac.cn}
\affiliation{Beijing National Laboratory for Condensed Matter Physics and Institute of Physics,
	Chinese Academy of Sciences, Beijing 100190, China}
\affiliation{Kavli Institute of Theoretical Sciences, University of Chinese Academy of Sciences,
	Beijing, 100190, China}
 \affiliation{New Cornerstone Science Laboratory, 
	Beijing, 100190, China}

\date{\today}

\begin{abstract}
Sublattice dichotomy in monolayer FeSe/SrTiO$_3$, signaling the breaking of symmetries exchanging the two Fe sublattices, has recently been reported. We propose that interband pairing serves as the origin of this dichotomy, regardless of whether the symmetry is broken in the normal state or in the pairing state. If symmetry breaking occurs in the normal state, the Fermi surfaces are sublattice-polarized, and the intersublattice $d$-wave pairing naturally acts as interband pairing, reproducing the observed dichotomy in the spectra. Alternatively, if symmetry breaking takes place in the pairing state, it manifests as the coexistence of intraband and interband pairing, with the constraint that interband pairings share the same sign while intraband pairings carry opposite signs. In both cases, interband pairing is indispensable, establishing it as a key ingredient for understanding superconductivity in monolayer FeSe/SrTiO$_3$.
\end{abstract}
\maketitle

\section{Introduction}

Iron-based superconductors have attracted sustained and widespread interest since their initial discovery \cite{Hirschfeld_2011, annurev:/content/journals/10.1146/annurev-conmatphys-020911-125055, Si2016,Fernandes2022, Kamihara2008, PhysRevLett.101.107006, doi:10.1073/pnas.0807325105, PhysRevB.78.060505, WANG2008538, PhysRevB.82.180520}, owing to their high critical temperatures, rich phase diagrams \cite{Fernandes2014, Fernandes_2017, Si2016, RevModPhys.87.855, Böhmer_2018, Fernandes2022}, and unconventional pairing mechanisms \cite{Hirschfeld_2011, annurev:/content/journals/10.1146/annurev-conmatphys-020911-125055, PhysRevLett.101.057003, PhysRevLett.101.087004, PhysRevLett.101.206404, PhysRevX.1.011009}.
Among these materials, a particularly striking development was the discovery of a record-high superconducting transition temperature ($\sim$ 65 K) in monolayer FeSe films grown on SrTiO$_3$ substrate\cite{FeSe/STO, He2013, Tan2013, Lee2014, ZHANG20151301, Ge2015,annurev:/content/journals/10.1146/annurev-conmatphys-031016-025242, Xu2021}. This system differs fundamentally from the usual iron pnictides. Angle-resolved photoemission spectroscopy (ARPES) measurements reveal that the hole-like bands near the $\Gamma$ point are below the Fermi level, leaving only electron pockets at the Brillouin zone corner \cite{Liu2012FeSeARPES}. The absence of the hole pockets invalidates the Fermi surface nesting scenario commonly invoked in the iron-based superconductors\cite{PhysRevLett.101.057003, PhysRevLett.101.087004}. Therefore, monolayer FeSe/SrTiO$_3$ has become a unique platform for exploring unconventional, interface-enhanced high-$T_c$ superconductivity\cite{Fan2015, PhysRevB.92.224514, PhysRevB.93.125129, PhysRevLett.117.077003, PhysRevLett.117.117001}.

\begin{figure}[H]
	\begin{center}
		\fig{3.4in}{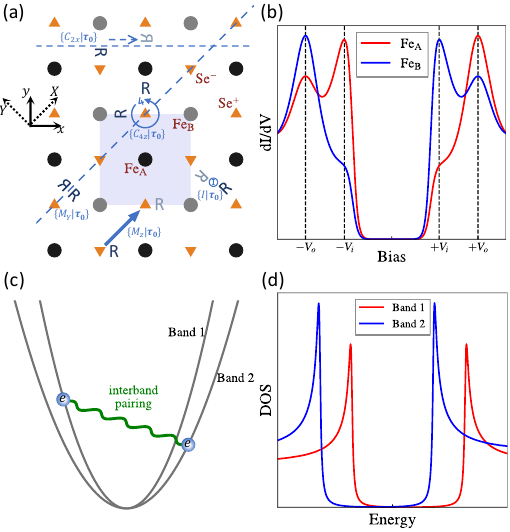}
		\caption{(a) (Left) Coordinate systems used in this paper. (Right) Top view of monolayer FeSe. Fe atoms are depicted with solid circles in black and gray, corresponding to the two Fe sublattices. Se atoms above and below the Fe plane are represented by upward-pointing and downward-pointing filled triangles, respectively. Fe$_\mathrm{A}$, Fe$_\mathrm{B}$, Se$^+$ and Se$^-$ are the notations for each type of atom. A single unit cell is indicated by the shaded square region. Symmetries exchanging the two Fe sublattices,  $\{I|\boldsymbol{\tau_0}\}$, $\{C_{4z}|\boldsymbol{\tau_0}\}$, $\{M_Y|\boldsymbol{\tau_0}\}$, $\{M_z|\boldsymbol{\tau_0}\}$, and $\{C_{2x}|\boldsymbol{\tau_0}\}$ where $\boldsymbol{\tau_0} = (1/2,1/2)$, are depicted in blue, with paired ``R" symbols visualizing the symmetry action---each pair represents the original object and its image under the corresponding operation. The dark blue ``R" is above the Fe plane, and the light blue one is below. The name of each symmetry is indicated nearby\cite{vafek_PhysRevB.88.134510}. (b) Schematic of the observed tunneling spectra on two sublattices. $V_i$ and $V_o$ are the inner and outer gaps. (c) Schematic of interband pairing. (d) Density of states (DOS) projected onto each band for a two-band superconductor with interband pairing\cite{PhysRevB.110.094517}. \label{fig1}}
	\end{center}
\end{figure}

The monolayer FeSe consists of a single layer of Fe atoms arranged in a square lattice, sandwiched by two Se atomic layers positioned alternately above (Se$^+$) and below (Se$^-$) the Fe plane (see Fig.~\ref{fig1}(a)). The alternating positioning of the Se atoms above and below the Fe plane doubles the size of the structural unit cell of the Fe square lattice, resulting in two Fe atoms per unit cell, named Fe$_\mathrm{A}$ and Fe$_\mathrm{B}$. For the pristine monolayer FeSe, these two sublattices are equivalent to each other due to symmetries, such as inversion symmetry whose center locates at the middle of the Fe$_\mathrm{A}$-Fe$_\mathrm{B}$ bond. In the material, however, the existence of substrate, electronic ordering, and many other issues may break the equivalence of the two sublattices.

Recently, sublattice dichotomy has been witnessed in the monolayer FeSe/SrTiO$_3$ in a scanning tunneling microscopy and spectroscopy (STM/STS) experiment \cite{ding2024sublattice}. Distinct dual tunneling spectra within the pairing gap corresponding to the two Fe sublattices (schematically illustrated in Fig.~\ref{fig1}(b)) are observed. Specifically, two superconducting gaps are identfied, namely $V_i$ and $V_o$. The coherence peak at $+V_i$ on $\mathrm{Fe_A}$ is lower 
\noindent
than that on $\mathrm{Fe_B}$, whereas the coherence peak at $-V_i$ on $\mathrm{Fe_A}$ is higher. Conversely, the intensity difference of the coherence peaks at $\pm V_o$ exhibits the opposite behavior. This intriguing phenomenon is reminiscent of interband pairing \cite{PhysRevB.110.094517, mei2025interband, PhysRevB.80.104507, PhysRevB.58.12307, Dolgov1987} in multiband superconductors. The interband pairing describes Cooper pairs formed by electrons from bands with different masses, as illustrated in Fig.~\ref{fig1}(c). Since the particle and hole sectors forming the Cooper pair are from the bands with distinct masses, such a subsystem intrinsically lacks particle–hole symmetry. Accordingly, the pairing between them necessarily produces particle–hole asymmetric spectra\cite{PhysRevB.110.094517}. Density of states (DOS) projected onto each band for a two-band superconductor with interband pairing are calculated in Ref.~\onlinecite{PhysRevB.110.094517} and also presented here in Fig.~\ref{fig1}(d). As shown, the coherence peaks of band~1 shift positively by half the energy difference between the two bands, while those of band~2 shift negatively by the same amount. Though the above analyses refer to the band-projected DOS in a two-band superconductor, the resulting spectra closely resemble the experimentally observed ones, i.e. the sublattice-projected DOS as displayed in Fig.~\ref{fig1}(b) and (d). Therefore, we propose that interband pairing underlies the sublattice dichotomy.


This paper is organized as follows. Section~\ref{symmetry} is devoted to the symmetry analysis, which identifies the symmetry requirement for realizing the sublattice dichotomy. Section~\ref{normalbreaking} and \ref{pairingbreaking} discuss two cases: symmetry breaking in the normal state and in the pairing state, both of which emphasize the role of interband pairing. Section~\ref{summary} is for summary and discussion.

\section{Symmetry analysis}\label{symmetry}

To figure out the origin of the sublattice dichotomy in monolayer FeSe/SrTiO$_3$, we start with a symmetry analysis. The pristine monolayer FeSe respects the space group $G = P4/nmm$, which is nonsymmorphic. Its quotient group $G/\mathcal{T}$ with $\mathcal{T}$ being the translation subgroup contains 16 symmetry operations as listed in Table.~\ref{tab:character}, the point group part of which is isomorphic to the point group $D_{4h}$. The symmetry operations in $G/\mathcal{T}$ can be classified into two categories; Half of the symmetry operations map the $\mathrm{Fe_A}$ sublattice to the $\mathrm{Fe_B}$ sublattice, and the others does not exchange the two Fe sublattices. We list the two categories of symmetry operations in Table.~\ref{tab:character}, and schematically illustrate the symmetries exchanging the two Fe sublattices in Fig.~\ref{fig1}(a). Here, we use the Seitz notation $\{g|\boldsymbol{\tau}\}$ where $g$ is a point group operation and $\boldsymbol{\tau}$ is a translation, to describe the symmetry operations.

Due to the symmetries exchanging the two Fe sublattices, it is natural to expect exactly the same experimental observations at $\mathrm{Fe_A}$ and $\mathrm{Fe_B}$ in the monolayer FeSe/SrTiO$_3$. Conversely, the sublattice dichotomy observed in recent experiments unambiguously indicates that all the symmetry operations mapping $\mathrm{Fe_A}$ to $\mathrm{Fe_B}$ are broken. To realize such a condition, orders belonging to different irreducible representations must be present. Notably, in the Nambu basis the BdG Hamiltonian of a superconductor always belongs to the $A_{1g}$ irreducible representation, if only one single pairing order is considered, regardless of the specific pairing symmetry. Hence, according to Table~\ref{tab:character} the problem reduces to identifying a $B_{2u}$ perturbation to the BdG Hamiltonian. Notice that the $B_{2u}$ perturbations can be the superconducting pairing orders or the orders in the normal bands.

\section{Normal state symmetry breaking} \label{normalbreaking}
We first consider the symmetries are broken in the normal state. We adopt the low-energy $k\cdot p$ model developed in Ref.~\onlinecite{vafek_PhysRevB.88.134510}, which resembles the electron pockets near the Brillouin zone corner, i.e. the M point, in the monolayer FeSe/SrTiO$_3$ \cite{ Liu2012FeSeARPES, ding2024sublattice}. The model respects the full symmetries of the space group $P4/nmm$, and takes the form $H_0 = \sum_{\mathbf{k},\sigma}\psi^\dagger_\sigma(\mathbf{k})h_M(\mathbf{k})\psi_\sigma(\mathbf{k})$ with
\begin{subequations}
    \begin{equation}
        h_M(\mathbf{k})=\begin{pmatrix}
            h^+_M(\mathbf{k}) & 0\\
            0 & h^-_M(\mathbf{k})
        \end{pmatrix},\label{eq:kp}
    \end{equation}
where
\begin{equation}
    h^\pm_M(\mathbf{k})=\begin{pmatrix}
        \epsilon_1+\frac{\mathbf{k}^2}{2m_1}\pm a_1k_xk_y & -iv_\pm(\mathbf{k}) \\
        iv_\pm(\mathbf{k}) & \epsilon_3+\frac{\mathbf{k}^2}{2m_3}\pm a_3k_xk_y
    \end{pmatrix},
\end{equation}
and 
\begin{equation}
    v_\pm(\mathbf{k})=v(\pm k_x+k_y)+p_1(\pm k_x^3+k_y^3) + p_2k_xk_y(k_x\pm k_y).
\end{equation}
\end{subequations}
The above Hamiltonian is written in the $d$-orbital basis 
\begin{equation}
    \psi_\sigma(\mathbf{k})=\frac{1}{\sqrt{2}}\begin{pmatrix}
        d^A_{Xz,\sigma}(\mathbf{k})-d^B_{Xz,\sigma}(\mathbf{k})\\
        d^A_{XY,\sigma}(\mathbf{k})+d^B_{XY,\sigma}(\mathbf{k})\\
        d^A_{Yz,\sigma}(\mathbf{k})+d^B_{Yz,\sigma}(\mathbf{k})\\
        d^A_{XY,\sigma}(\mathbf{k})-d^B_{XY,\sigma}(\mathbf{k})
    \end{pmatrix},\label{eq:basis}
\end{equation}
where $\mathbf{k}$ is measured from the M point, $\sigma$ is the spin index, and $A$ ($B$) stands for the the $\mathrm{Fe_A}$ ($\mathrm{Fe_B}$) sublattice. 

\begin{figure}[t]
	\begin{center}
		\fig{3.4in}{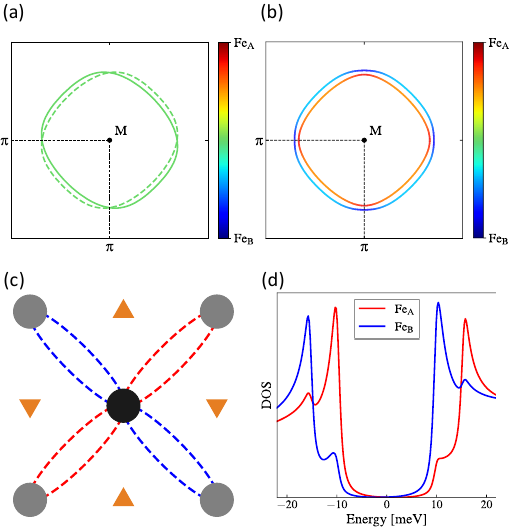}
		\caption{(a) Fermi surfaces without symmetry-breaking terms. Parameters are taken as $\epsilon_1=-95.00$, $\epsilon_3=-60.00$, $\frac{1}{2m_1}=-2.40$, $\frac{1}{2m_3}=20.58$, $a_1=38.06$, $a_3=-48.56$, $v=39.90$, $p_1=-0.63$, $p_2=-1.87$ in unit of meV to mimic the experimental Fermi surface results \cite{ding2024sublattice}. (b) Fermi surfaces with symmetry-breaking terms. Parameters in $h'_M(\mathbf{k})$ are taken as $v_1=v_2=v_3=v_4=1.00$~meV. The color in (a) and (b) represents the sublattice weight. (c) Schematic illustration of nodeless $d$-wave pairing in real space. The pairing is positive (red) in the $X-$direction and negative (blue) in the $Y-$direction. (d) Calculated DOS with the combination of normal state symmetry breaking Eq.~\eqref{eq:perturbation_band} and nodeless $d$-wave pairing Eq.~\eqref{eq:b2g}, where $\Delta_d = 12.20$ meV.\label{fig2}}
	\end{center}
\end{figure}

We use the above model to fit the experimentally reported Fermi surfaces of the monolayer FeSe/SrTiO$_3$\cite{ding2024sublattice}, and derive the parameters as $\epsilon_1=-95.00$, $\epsilon_3=-60.00$, $\frac{1}{2m_1}=-2.40$, $\frac{1}{2m_3}=20.58$, $a_1=38.06$, $a_3=-48.56$, $v=39.90$, $p_1=-0.63$ and $p_2=-1.87$, which are all in unit of meV. The corresponding Fermi surfaces showing the sublattice weight are presented in Fig.~\ref{fig2}(a). As shown in the figure, on the Fermi surfaces the two sublattices are evenly mixed. This arises from the constraint imposed by the glide-mirror symmetry $\{M_z|\boldsymbol{\tau_0}\}$ which exchanges the two Fe sublattices but leaves $\mathbf{k}$ invariant.

\begin{widetext}
\begin{center}
\begin{table}[t]
\centering
\caption{Character table of $G/\mathcal{T}$ (isomorphic to $D_{4h}$), where $\boldsymbol{\tau_0}$ stands for the translation $(1/2,1/2)$. The symmetry operations exchanging the two Fe sublattices are marked in blue.  \label{tab:character}}
\begin{tabular}{ccccccccccc}
\hhline{===========}
\\[-0.9em]
& $\{E|\mathbf{0}\}$ & 2$\{S_{4z}|\mathbf{0}\}$ & $\{C_{2z}|\mathbf{0}\}$ & $\{C_{2\,X/Y}|\mathbf{0}\}$ & $\{M_{x/y}|\mathbf{0}\}$ & \textcolor{blue}{$\{I|\boldsymbol{\tau_0}\}$} & \textcolor{blue}{2$\{C_{4z}|\boldsymbol{\tau_0}\}$} & \textcolor{blue}{$\{M_{X/Y}|\boldsymbol{\tau_0}\}$} & \textcolor{blue}{$\{M_z|\boldsymbol{\tau_0}\}$} & \textcolor{blue}{$\{C_{2\,x/y}|\boldsymbol{\tau_0}\}$}\\[0.1em]
\hhline{-----------}
\\[-0.9em]
$A_{1g}$ & 1 & 1 & 1 & 1 & 1 & 1 & 1 & 1 & 1 & 1\\
$A_{2g}$ & 1 & 1 & 1 & -1 & -1 & 1 & 1 & -1 & 1 & -1 \\
$B_{1g}$ & 1 & -1 & 1 & -1 & 1 & 1 & -1 & -1 & 1 & 1\\
$B_{2g}$ & 1 & -1 & 1 & 1 & -1 & 1 & -1 & 1 & 1 & -1\\
$E_g$ & 2 & 0 & -2 & 0 & 0 & 2 & 0 & 0 & -2 & 0  \\
$A_{1u}$ & 1 & -1 & 1 & 1 & -1 & -1 & 1 & -1 & -1 & 1 \\
$A_{2u}$ & 1 & -1 & 1 & -1 & 1 & -1 & 1 & 1 & -1 & -1 \\
$B_{1u}$ & 1 & 1 & 1 & -1 & -1 & -1 & -1 & 1 & -1 & 1 \\
$B_{2u}$ & 1 & 1 & 1 & 1 & 1 & -1 & -1 & -1 & -1 & -1 \\
$E_u$ & 2 & 0 & -2 & 0 & 0 & -2 & 0 & 0 & 2 & 0 \\[0.1em]
\hhline{===========}
\end{tabular}
\end{table}
\end{center}
\end{widetext}

Based on the low-energy effective model, we now consider perturbations that break all the symmetries exchanging the two Fe sublattices, i.e. the $B_{2u}$ perturbations in the normal bands. Note that the $B_{2u}$ perturbations in the normal state must also be the $B_{2u}$ perturbations in the superconducting state. Under the basis in Eq.~\eqref{eq:basis}, up to the $\mathbf{k}^2$ order the perturbations take the form
\begin{equation}
\small
    h'_M(\mathbf{k})=\begin{pmatrix}
        0 & 0 & v_3(k_x^2{-}k_y^2) & -iv_4(k_x{+}k_y) \\
        0 & 0 & i v_4({-}k_x{+}k_y) & v_1 {+} v_2\mathbf{k}^2\\
        v_3(k_x^2{-}k_y^2) & {-}i v_4({-}k_x{+}k_y) & 0 & 0\\
        i v_4(k_x{+}k_y) & v_1 {+} v_2\mathbf{k}^2 & 0 & 0
    \end{pmatrix}.\label{eq:perturbation_band}
\end{equation}
Upon adding $h'_M$ into $h_M$ with $v_1=v_2=v_3=v_4=1.00$~meV, the corresponding Fermi surfaces are shown in Fig.~\ref{fig2}(b). Clearly, the sublattice weights are clearly redistributed; Specifically, the inner Fermi surface is predominantly contributed by $\mathrm{Fe_A}$, while the outer one is mainly from $\mathrm{Fe_B}$. The results suggest a correspondence between the sublattices and the bands. Thus, the intersublattice pairing effectively plays the role of the interband pairing, potentially giving rise to the features of Fig.~\ref{fig1}(d). In the intersublattice pairing channel, the $s$-wave pairing gives nodal superconducting state. Considering the nodeless superconducting gap observed in the monolayer FeSe/SrTiO$_3$, we focus on the intersublattice $d$-wave pairing\cite{PhysRevLett.119.267001, PhysRevB.83.100515, PhysRevB.84.024529, PhysRevB.88.094522}, more specifically the $d$-wave pairing between the nearest-neighbor $\mathrm{Fe_A}$ and $\mathrm{Fe_B}$ sites. The $d$-wave pairing order changes sign under the $S_4$ symmetry as schematically illustrated in Fig.~\ref{fig2}(c). As the Fermi surfaces are located around $M$ which is $S_4$ invariant, superconducting nodes should have existed on the Fermi surfaces; However, due to the interband pairing nature of the intersublattice pairing, the $d$-wave pairing state is actually fully gapped.

The leading order of the nodeless $d$-wave pairing, expressed in form of $\psi^\dagger_\uparrow(\mathbf{k})\Delta(\mathbf{k})[\psi^\dagger_\downarrow(\mathbf{-k})]^T$, is given by
\begin{equation}
    \Delta(\mathbf{k})=\Delta_d\begin{pmatrix}
        1 & 0 & 0 & 0 \\
        0 & 1 & 0 & 0 \\
        0 & 0 & -1 & 0 \\
        0 & 0 & 0 & -1
    \end{pmatrix}.\label{eq:b2g}
\end{equation}
Combining the normal-state perturbations in Eq.\eqref{eq:perturbation_band} and the nodeless $d$-wave pairing in Eq.\eqref{eq:b2g}, we calculate the DOS at $\mathrm{Fe_A}$ and $\mathrm{Fe_B}$, and the results are presented in Fig.~\ref{fig2}(d). As expected, the sublattice dichotomy of the superconducting coherence peaks is realized, due to the sublattice-polarized Fermi surfaces and the interband nature of the $d$-wave pairing. 

Before ending this part, we note that incorporating the $s_\pm$ pairing component, alongside the above $d$-wave pairing, does not necessarily alter the sublattice dichotomy, provided that the magnitude of $s_\pm$ component is not dominant.

\section{Pairing state symmetry breaking}\label{pairingbreaking}

The sublattice dichotomy in the monolayer FeSe/SrTiO$_3$ may also arise from the mix of the different pairing orders. To break the symmetries mapping $\mathrm{Fe_A}$ to $\mathrm{Fe_B}$, according to Table.~\ref{tab:character} one need to consider the following pairing combinations, $A_{1g}+B_{2u}$, $A_{2g}+B_{1u}$, $B_{1g}+A_{2u}$ and $B_{2g}+A_{1u}$. Here, we focus on the pairings belonging to the 1D irreducible representations. In fact, for each of the above pairing combinations, if we consider the even-parity (odd-parity) order separately the system is invariant under the crystalline symmetries, i.e. the BdG Hamiltonian belonging to the $A_{1g}$ irreducible representation, and the odd-parity (even-parity) order serves as the $B_{2u}$ perturbations in the sense of the BdG Hamiltonian.


For the iron-based superconductors, it has been revealed \cite{PhysRevX.3.031004} that for the 1D irreducible representations the even-parity pairing orders correspond to intraband pairing, whereas the odd-parity orders correspond to interband pairing.
To obtain the sublattice dichotomy in the monolayer FeSe/SrTiO$_3$, the coexistence of intraband and interband pairing serves as the starting point, and additional constraints on their relative phases must be imposed as we shall show in the following.

Before considering the monolayer FeSe/SrTiO$_3$, it is helpful to utilize a simple two-band model to develop some insights. We denote the band operators as $c^\dagger_{1k\sigma}$ and $c^\dagger_{2k\sigma}$, and assume simple dispersions $\gamma_{1}k^2-\mu$ and $\gamma_{2}k^2-\mu$ ($\gamma_1>\gamma_2$) for the two bands. The BdG Hamiltonian under the basis $\Psi_k^\dagger=(c^\dagger_{1k\uparrow}, c^\dagger_{2k\uparrow}, c_{1,-k\downarrow}, c_{2,-k\downarrow})$ is
\begin{equation}
    H_{\mathrm{BdG}}=\begin{pmatrix}
        \gamma_1 k^2{-}\mu & 0 & \Delta_{a1} & \Delta_{b1} \\
        0 & \gamma_2 k^2{-}\mu & \Delta_{b2} & \Delta_{a2} \\
        \Delta^*_{a1} & \Delta^*_{b2} & -\gamma_1k^2{+}\mu & 0 \\
        \Delta^*_{b1} & \Delta^*_{a2} & 0 & -\gamma_2k^2{+}\mu
    \end{pmatrix},\label{eq:BdG}
\end{equation}
where $\Delta_{a1(2)}$ and $\Delta_{b1(2)}$ are the intraband and interband pairings, respectively. To determine the conditions under which the sublattice dichotomy emerges, we perform a perturbation analysis (details in the Supplemental Material) and find that the only pairing configuration yielding the sublattice dichotomy is $\Delta_{b1}=\Delta_{b2}$ and $\Delta_{a1}=-\Delta_{a2}$, i.e. the interband pairing having the same sign and the intraband pairing having opposite signs.

\begin{figure}[t]
	\begin{center}
		\fig{3.4in}{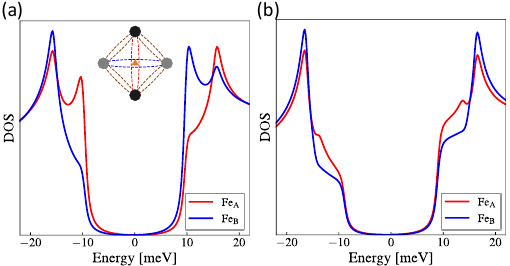}
		\caption{Calculated DOSs with pairing combinations of (a) Eq.~\eqref{eq:b2u} with \eqref{eq:a1g2}, (b) \eqref{eq:b2u} with \eqref{eq:a1g0}. Parameters are (a) $\Delta_{b2u}=15$ meV, $\Delta_{a1g}=10$ meV, (b) $\Delta_{b2u}=18$ meV, $\Delta_{a1g}=-0.5$ meV. The inset in (a) is the schematic illustration of the pairing combination of Eq.~\eqref{eq:b2u} with \eqref{eq:a1g2} in real space. The intrasublattice pairing is positive (red) on $\mathrm{Fe_A}$–$\mathrm{Fe_A}$ bonds and negative (blue) on $\mathrm{Fe_B}$–$\mathrm{Fe_B}$ bonds, while all intersublattice pairings share the same sign (brown). \label{fig3}}
	\end{center}
\end{figure}

We examine the above criterion in the monolayer FeSe/SrTiO$_3$. Specifically, we consider the pairings $A_{1g}+B_{2u}$ as an example in the $k\cdot p$ model in Eq.~\eqref{eq:kp}. The leading order of the $B_{2u}$ pairing under the basis in Eq.~\eqref{eq:basis} is 
\begin{equation}
    B_{2u}: \Delta(\mathbf{k})=\Delta_{b2u}\begin{pmatrix}
        0 & 0 & 0 & 0 \\
        0 & 0 & 0 & 1 \\
        0 & 0 & 0 & 0 \\
        0 & 1 & 0 & 0
    \end{pmatrix}.\label{eq:b2u}
\end{equation}
Based on the $k \cdot p$ in Eq.~\eqref{eq:kp}, it can be found the $B_{2u}$ pairing in Eq.\eqref{eq:b2u} is actually the pure interband pairing which occurs in the $d_{XY}$ orbital. Moreover, the $B_{2u}$ pairing in Eq.\eqref{eq:b2u} is even with respect to the band indices, satisfying the criterion for the sublattice dichotomy. For the $A_{1g}$ pairing, the two leading orders preserved up to $\mathbf{k}^2$ take the form
\begin{subequations}
    \begin{equation}
        A_{1g}: \Delta(\mathbf{k})=\Delta_{a1g}\begin{pmatrix}
            1 & 0 & 0 & 0 \\
            0 & 1 & 0 & 0 \\
            0 & 0 & 1 & 0 \\
            0 & 0 & 0 & 1
        \end{pmatrix},\label{eq:a1g0}
    \end{equation}
    \begin{equation}
        A_{1g}: \Delta(\mathbf{k})=\Delta_{a1g}\begin{pmatrix}
            k_xk_y & 0 & 0 & 0 \\
            0 & k_xk_y & 0 & 0 \\
            0 & 0 & -k_xk_y & 0 \\
            0 & 0 & 0 & -k_xk_y
        \end{pmatrix}.\label{eq:a1g2}
    \end{equation}
\end{subequations}
Based on the criterion developed in the above, only the $A_{1g}$ pairing in Eq.\eqref{eq:a1g2} which is odd with respect to the band indices satisfies the condition. We further simulate the superconducting DOSs with the $B_{2u}$ pairing in Eq.~\eqref{eq:b2u} and the $A_{1g}$ pairing in Eq.\eqref{eq:a1g2}, and present the results in Fig.~\ref{fig3}(a). As expected, the sublattice dichotomy is observed at the $\mathrm{Fe_A}$ and $\mathrm{Fe_B}$ sites, consistent with the experiments. To compare, we also calculate the case with the $B_{2u}$ pairing in Eq.~\eqref{eq:b2u} and the $A_{1g}$ pairing in Eq.\eqref{eq:a1g0}; As shown in Fig.~\ref{fig3}(b), no sublattice dichotomy appears.

In real space, for the above pairing combination responsible for the sublattice dichotomy, the $B_{2u}$ order in Eq.~\eqref{eq:b2u} corresponds to the nearest-neighbour intrasublattice pairing which is $S_4$ invariant but have opposite signs in the $\mathrm{Fe_A}$ sublattice and $\mathrm{Fe_B}$ sublattice; While the $A_{1g}$ order in Eq.\eqref{eq:a1g2} describes the nearest-neighbour intersublattice pairing with a uniform pairing sign. We schematically illustrate the two pairing orders in the inset of Fig.~\ref{fig3}(a). Such a pairing combination has been proposed in Ref.~\onlinecite{PhysRevX.3.031004}.

\section{Summary and discussion}\label{summary}
In summary, we have identified interband pairing as a key ingredient in understanding the sublattice dichotomy observed in the monolayer FeSe/SrTiO$_3$\cite{ding2024sublattice}. Two possibilities are recognized for such symmetry-breaking. In the first case, symmetry-breaking is present in the normal state, leading to sublattice-polarized Fermi surfaces. The intersublattice $d$-wave pairing effectively acts as interband pairing, which can naturally reproduce the experimentally observed dichotomy in the spectra. 
However, we notice the nodeless $d$-wave pairing, is challenged by vortex-core STM/STS measurements\cite{PhysRevLett.124.097001}, which report a CdGM spectrum inconsistent with $d$-wave–based theoretical predictions\cite{Xiang2025Accidental}. This discrepancy calls further explorations.
In the second case, symmetry breaking emerges at the superconducting level through a mixing of both intraband and interband pairing. Here, an additional constraint is required: while the interband pairings must share the same sign, the intraband pairings must carry opposite signs.

\textit{Note added.--} During the preparation of this manuscript, we became aware of Ref.~\onlinecite{roig2025origin} , which also investigates the sublattice dichotomy in monolayer FeSe/SrTiO$_3$.

\section{acknowledgments}
We acknowledge the support by the Ministry of Science and Technology  (Grant No. 2022YFA1403900), the National Natural Science Foundation of China (Grant NSFC-12494594, No. NSFC-12174428, Grant No. NSFC-12304163), the New Cornerstone Investigator Program, the Chinese Academy of Sciences Project for Young Scientists in Basic Research (2022YSBR-048), and the Beijing Institute of Technology Research Fund Program for Young Scholars.

\bibliography{reference}

\begin{thebibliography}{47}%
\makeatletter
\providecommand \@ifxundefined [1]{%
 \@ifx{#1\undefined}
}%
\providecommand \@ifnum [1]{%
 \ifnum #1\expandafter \@firstoftwo
 \else \expandafter \@secondoftwo
 \fi
}%
\providecommand \@ifx [1]{%
 \ifx #1\expandafter \@firstoftwo
 \else \expandafter \@secondoftwo
 \fi
}%
\providecommand \natexlab [1]{#1}%
\providecommand \enquote  [1]{``#1''}%
\providecommand \bibnamefont  [1]{#1}%
\providecommand \bibfnamefont [1]{#1}%
\providecommand \citenamefont [1]{#1}%
\providecommand \href@noop [0]{\@secondoftwo}%
\providecommand \href [0]{\begingroup \@sanitize@url \@href}%
\providecommand \@href[1]{\@@startlink{#1}\@@href}%
\providecommand \@@href[1]{\endgroup#1\@@endlink}%
\providecommand \@sanitize@url [0]{\catcode `\\12\catcode `\$12\catcode
  `\&12\catcode `\#12\catcode `\^12\catcode `\_12\catcode `\%12\relax}%
\providecommand \@@startlink[1]{}%
\providecommand \@@endlink[0]{}%
\providecommand \url  [0]{\begingroup\@sanitize@url \@url }%
\providecommand \@url [1]{\endgroup\@href {#1}{\urlprefix }}%
\providecommand \urlprefix  [0]{URL }%
\providecommand \Eprint [0]{\href }%
\providecommand \doibase [0]{http://dx.doi.org/}%
\providecommand \selectlanguage [0]{\@gobble}%
\providecommand \bibinfo  [0]{\@secondoftwo}%
\providecommand \bibfield  [0]{\@secondoftwo}%
\providecommand \translation [1]{[#1]}%
\providecommand \BibitemOpen [0]{}%
\providecommand \bibitemStop [0]{}%
\providecommand \bibitemNoStop [0]{.\EOS\space}%
\providecommand \EOS [0]{\spacefactor3000\relax}%
\providecommand \BibitemShut  [1]{\csname bibitem#1\endcsname}%
\let\auto@bib@innerbib\@empty
\bibitem [{\citenamefont {Hirschfeld}\ \emph {et~al.}(2011)\citenamefont
  {Hirschfeld}, \citenamefont {Korshunov},\ and\ \citenamefont
  {Mazin}}]{Hirschfeld_2011}%
  \BibitemOpen
  \bibfield  {author} {\bibinfo {author} {\bibfnamefont {P~J}\ \bibnamefont
  {Hirschfeld}}, \bibinfo {author} {\bibfnamefont {M~M}\ \bibnamefont
  {Korshunov}}, \ and\ \bibinfo {author} {\bibfnamefont {I~I}\ \bibnamefont
  {Mazin}},\ }\bibfield  {title} {\enquote {\bibinfo {title} {Gap symmetry and
  structure of {Fe}-based superconductors},}\ }\href {\doibase
  10.1088/0034-4885/74/12/124508} {\bibfield  {journal} {\bibinfo  {journal}
  {Reports on Progress in Physics}\ }\textbf {\bibinfo {volume} {74}},\
  \bibinfo {pages} {124508} (\bibinfo {year} {2011})}\BibitemShut {NoStop}%
\bibitem [{\citenamefont
  {Chubukov}(2012)}]{annurev:/content/journals/10.1146/annurev-conmatphys-020911-125055}%
  \BibitemOpen
  \bibfield  {author} {\bibinfo {author} {\bibfnamefont {Andrey}\ \bibnamefont
  {Chubukov}},\ }\bibfield  {title} {\enquote {\bibinfo {title} {Pairing
  mechanism in {Fe}-based superconductors},}\ }\href {\doibase
  https://doi.org/10.1146/annurev-conmatphys-020911-125055} {\bibfield
  {journal} {\bibinfo  {journal} {Annual Review of Condensed Matter Physics}\
  }\textbf {\bibinfo {volume} {3}},\ \bibinfo {pages} {57--92} (\bibinfo {year}
  {2012})}\BibitemShut {NoStop}%
\bibitem [{\citenamefont {Si}\ \emph {et~al.}(2016)\citenamefont {Si},
  \citenamefont {Yu},\ and\ \citenamefont {Abrahams}}]{Si2016}%
  \BibitemOpen
  \bibfield  {author} {\bibinfo {author} {\bibfnamefont {Qimiao}\ \bibnamefont
  {Si}}, \bibinfo {author} {\bibfnamefont {Rong}\ \bibnamefont {Yu}}, \ and\
  \bibinfo {author} {\bibfnamefont {Elihu}\ \bibnamefont {Abrahams}},\
  }\bibfield  {title} {\enquote {\bibinfo {title} {High-temperature
  superconductivity in iron pnictides and chalcogenides},}\ }\href {\doibase
  10.1038/natrevmats.2016.17} {\bibfield  {journal} {\bibinfo  {journal}
  {Nature Reviews Materials}\ }\textbf {\bibinfo {volume} {1}},\ \bibinfo
  {pages} {16017} (\bibinfo {year} {2016})}\BibitemShut {NoStop}%
\bibitem [{\citenamefont {Fernandes}\ \emph {et~al.}(2022)\citenamefont
  {Fernandes}, \citenamefont {Coldea}, \citenamefont {Ding}, \citenamefont
  {Fisher}, \citenamefont {Hirschfeld},\ and\ \citenamefont
  {Kotliar}}]{Fernandes2022}%
  \BibitemOpen
  \bibfield  {author} {\bibinfo {author} {\bibfnamefont {Rafael~M.}\
  \bibnamefont {Fernandes}}, \bibinfo {author} {\bibfnamefont {Amalia~I.}\
  \bibnamefont {Coldea}}, \bibinfo {author} {\bibfnamefont {Hong}\ \bibnamefont
  {Ding}}, \bibinfo {author} {\bibfnamefont {Ian~R.}\ \bibnamefont {Fisher}},
  \bibinfo {author} {\bibfnamefont {P.~J.}\ \bibnamefont {Hirschfeld}}, \ and\
  \bibinfo {author} {\bibfnamefont {Gabriel}\ \bibnamefont {Kotliar}},\
  }\bibfield  {title} {\enquote {\bibinfo {title} {Iron pnictides and
  chalcogenides: a new paradigm for superconductivity},}\ }\href {\doibase
  10.1038/s41586-021-04073-2} {\bibfield  {journal} {\bibinfo  {journal}
  {Nature}\ }\textbf {\bibinfo {volume} {601}},\ \bibinfo {pages} {35--44}
  (\bibinfo {year} {2022})}\BibitemShut {NoStop}%
\bibitem [{\citenamefont {Kamihara}\ \emph {et~al.}(2008)\citenamefont
  {Kamihara}, \citenamefont {Watanabe}, \citenamefont {Hirano},\ and\
  \citenamefont {Hosono}}]{Kamihara2008}%
  \BibitemOpen
  \bibfield  {author} {\bibinfo {author} {\bibfnamefont {Yoichi}\ \bibnamefont
  {Kamihara}}, \bibinfo {author} {\bibfnamefont {Takumi}\ \bibnamefont
  {Watanabe}}, \bibinfo {author} {\bibfnamefont {Masahiro}\ \bibnamefont
  {Hirano}}, \ and\ \bibinfo {author} {\bibfnamefont {Hideo}\ \bibnamefont
  {Hosono}},\ }\bibfield  {title} {\enquote {\bibinfo {title} {Iron-based
  layered superconductor {La[O$_{1-x}$F$_x$]FeAs} ($x$ = 0.05--0.12) with
  {T}$_c$ = 26 {K}},}\ }\href {\doibase 10.1021/ja800073m} {\bibfield
  {journal} {\bibinfo  {journal} {Journal of the American Chemical Society}\
  }\textbf {\bibinfo {volume} {130}},\ \bibinfo {pages} {3296--3297} (\bibinfo
  {year} {2008})}\BibitemShut {NoStop}%
\bibitem [{\citenamefont {Rotter}\ \emph {et~al.}(2008)\citenamefont {Rotter},
  \citenamefont {Tegel},\ and\ \citenamefont
  {Johrendt}}]{PhysRevLett.101.107006}%
  \BibitemOpen
  \bibfield  {author} {\bibinfo {author} {\bibfnamefont {Marianne}\
  \bibnamefont {Rotter}}, \bibinfo {author} {\bibfnamefont {Marcus}\
  \bibnamefont {Tegel}}, \ and\ \bibinfo {author} {\bibfnamefont {Dirk}\
  \bibnamefont {Johrendt}},\ }\bibfield  {title} {\enquote {\bibinfo {title}
  {Superconductivity at 38 {K} in the iron arsenide
  {$({\mathrm{Ba}}_{1\ensuremath{-}x}{\mathrm{K}}_{x}){\mathrm{Fe}}_{2}{\mathrm{As}}_{2}$}},}\
  }\href {\doibase 10.1103/PhysRevLett.101.107006} {\bibfield  {journal}
  {\bibinfo  {journal} {Phys. Rev. Lett.}\ }\textbf {\bibinfo {volume} {101}},\
  \bibinfo {pages} {107006} (\bibinfo {year} {2008})}\BibitemShut {NoStop}%
\bibitem [{\citenamefont {Hsu}\ \emph {et~al.}(2008)\citenamefont {Hsu},
  \citenamefont {Luo}, \citenamefont {Yeh}, \citenamefont {Chen}, \citenamefont
  {Huang}, \citenamefont {Wu}, \citenamefont {Lee}, \citenamefont {Huang},
  \citenamefont {Chu}, \citenamefont {Yan},\ and\ \citenamefont
  {Wu}}]{doi:10.1073/pnas.0807325105}%
  \BibitemOpen
  \bibfield  {author} {\bibinfo {author} {\bibfnamefont {Fong-Chi}\
  \bibnamefont {Hsu}}, \bibinfo {author} {\bibfnamefont {Jiu-Yong}\
  \bibnamefont {Luo}}, \bibinfo {author} {\bibfnamefont {Kuo-Wei}\ \bibnamefont
  {Yeh}}, \bibinfo {author} {\bibfnamefont {Ta-Kun}\ \bibnamefont {Chen}},
  \bibinfo {author} {\bibfnamefont {Tzu-Wen}\ \bibnamefont {Huang}}, \bibinfo
  {author} {\bibfnamefont {Phillip~M.}\ \bibnamefont {Wu}}, \bibinfo {author}
  {\bibfnamefont {Yong-Chi}\ \bibnamefont {Lee}}, \bibinfo {author}
  {\bibfnamefont {Yi-Lin}\ \bibnamefont {Huang}}, \bibinfo {author}
  {\bibfnamefont {Yan-Yi}\ \bibnamefont {Chu}}, \bibinfo {author}
  {\bibfnamefont {Der-Chung}\ \bibnamefont {Yan}}, \ and\ \bibinfo {author}
  {\bibfnamefont {Maw-Kuen}\ \bibnamefont {Wu}},\ }\bibfield  {title} {\enquote
  {\bibinfo {title} {Superconductivity in the {PbO}-type structure
  $\alpha$-{FeSe}},}\ }\href {\doibase 10.1073/pnas.0807325105} {\bibfield
  {journal} {\bibinfo  {journal} {Proceedings of the National Academy of
  Sciences}\ }\textbf {\bibinfo {volume} {105}},\ \bibinfo {pages}
  {14262--14264} (\bibinfo {year} {2008})}\BibitemShut {NoStop}%
\bibitem [{\citenamefont {Tapp}\ \emph {et~al.}(2008)\citenamefont {Tapp},
  \citenamefont {Tang}, \citenamefont {Lv}, \citenamefont {Sasmal},
  \citenamefont {Lorenz}, \citenamefont {Chu},\ and\ \citenamefont
  {Guloy}}]{PhysRevB.78.060505}%
  \BibitemOpen
  \bibfield  {author} {\bibinfo {author} {\bibfnamefont {Joshua~H.}\
  \bibnamefont {Tapp}}, \bibinfo {author} {\bibfnamefont {Zhongjia}\
  \bibnamefont {Tang}}, \bibinfo {author} {\bibfnamefont {Bing}\ \bibnamefont
  {Lv}}, \bibinfo {author} {\bibfnamefont {Kalyan}\ \bibnamefont {Sasmal}},
  \bibinfo {author} {\bibfnamefont {Bernd}\ \bibnamefont {Lorenz}}, \bibinfo
  {author} {\bibfnamefont {Paul C.~W.}\ \bibnamefont {Chu}}, \ and\ \bibinfo
  {author} {\bibfnamefont {Arnold~M.}\ \bibnamefont {Guloy}},\ }\bibfield
  {title} {\enquote {\bibinfo {title} {{LiFeAs}: {An} intrinsic {FeAs}-based
  superconductor with {${T}_{c}=18\text{ }\text{K}$}},}\ }\href {\doibase
  10.1103/PhysRevB.78.060505} {\bibfield  {journal} {\bibinfo  {journal} {Phys.
  Rev. B}\ }\textbf {\bibinfo {volume} {78}},\ \bibinfo {pages} {060505}
  (\bibinfo {year} {2008})}\BibitemShut {NoStop}%
\bibitem [{\citenamefont {Wang}\ \emph {et~al.}(2008)\citenamefont {Wang},
  \citenamefont {Liu}, \citenamefont {Lv}, \citenamefont {Gao}, \citenamefont
  {Yang}, \citenamefont {Yu}, \citenamefont {Li},\ and\ \citenamefont
  {Jin}}]{WANG2008538}%
  \BibitemOpen
  \bibfield  {author} {\bibinfo {author} {\bibfnamefont {X.C.}\ \bibnamefont
  {Wang}}, \bibinfo {author} {\bibfnamefont {Q.Q.}\ \bibnamefont {Liu}},
  \bibinfo {author} {\bibfnamefont {Y.X.}\ \bibnamefont {Lv}}, \bibinfo
  {author} {\bibfnamefont {W.B.}\ \bibnamefont {Gao}}, \bibinfo {author}
  {\bibfnamefont {L.X.}\ \bibnamefont {Yang}}, \bibinfo {author} {\bibfnamefont
  {R.C.}\ \bibnamefont {Yu}}, \bibinfo {author} {\bibfnamefont {F.Y.}\
  \bibnamefont {Li}}, \ and\ \bibinfo {author} {\bibfnamefont {C.Q.}\
  \bibnamefont {Jin}},\ }\bibfield  {title} {\enquote {\bibinfo {title} {The
  superconductivity at 18 {K} in {LiFeAs} system},}\ }\href {\doibase
  https://doi.org/10.1016/j.ssc.2008.09.057} {\bibfield  {journal} {\bibinfo
  {journal} {Solid State Communications}\ }\textbf {\bibinfo {volume} {148}},\
  \bibinfo {pages} {538--540} (\bibinfo {year} {2008})}\BibitemShut {NoStop}%
\bibitem [{\citenamefont {Guo}\ \emph {et~al.}(2010)\citenamefont {Guo},
  \citenamefont {Jin}, \citenamefont {Wang}, \citenamefont {Wang},
  \citenamefont {Zhu}, \citenamefont {Zhou}, \citenamefont {He},\ and\
  \citenamefont {Chen}}]{PhysRevB.82.180520}%
  \BibitemOpen
  \bibfield  {author} {\bibinfo {author} {\bibfnamefont {Jiangang}\
  \bibnamefont {Guo}}, \bibinfo {author} {\bibfnamefont {Shifeng}\ \bibnamefont
  {Jin}}, \bibinfo {author} {\bibfnamefont {Gang}\ \bibnamefont {Wang}},
  \bibinfo {author} {\bibfnamefont {Shunchong}\ \bibnamefont {Wang}}, \bibinfo
  {author} {\bibfnamefont {Kaixing}\ \bibnamefont {Zhu}}, \bibinfo {author}
  {\bibfnamefont {Tingting}\ \bibnamefont {Zhou}}, \bibinfo {author}
  {\bibfnamefont {Meng}\ \bibnamefont {He}}, \ and\ \bibinfo {author}
  {\bibfnamefont {Xiaolong}\ \bibnamefont {Chen}},\ }\bibfield  {title}
  {\enquote {\bibinfo {title} {Superconductivity in the iron selenide
  {${\text{K}}_{x}{\text{Fe}}_{2}{\text{Se}}_{2}$}
  $(0\ensuremath{\le}x\ensuremath{\le}1.0)$},}\ }\href {\doibase
  10.1103/PhysRevB.82.180520} {\bibfield  {journal} {\bibinfo  {journal} {Phys.
  Rev. B}\ }\textbf {\bibinfo {volume} {82}},\ \bibinfo {pages} {180520}
  (\bibinfo {year} {2010})}\BibitemShut {NoStop}%
\bibitem [{\citenamefont {Fernandes}\ \emph {et~al.}(2014)\citenamefont
  {Fernandes}, \citenamefont {Chubukov},\ and\ \citenamefont
  {Schmalian}}]{Fernandes2014}%
  \BibitemOpen
  \bibfield  {author} {\bibinfo {author} {\bibfnamefont {Rafael~M.}\
  \bibnamefont {Fernandes}}, \bibinfo {author} {\bibfnamefont {Andrey~V.}\
  \bibnamefont {Chubukov}}, \ and\ \bibinfo {author} {\bibfnamefont {J{\"o}rg}\
  \bibnamefont {Schmalian}},\ }\bibfield  {title} {\enquote {\bibinfo {title}
  {What drives nematic order in iron-based superconductors?}}\ }\href {\doibase
  10.1038/nphys2877} {\bibfield  {journal} {\bibinfo  {journal} {Nature
  Physics}\ }\textbf {\bibinfo {volume} {10}},\ \bibinfo {pages} {97--104}
  (\bibinfo {year} {2014})}\BibitemShut {NoStop}%
\bibitem [{\citenamefont {Fernandes}\ and\ \citenamefont
  {Chubukov}(2016)}]{Fernandes_2017}%
  \BibitemOpen
  \bibfield  {author} {\bibinfo {author} {\bibfnamefont {Rafael~M}\
  \bibnamefont {Fernandes}}\ and\ \bibinfo {author} {\bibfnamefont {Andrey~V}\
  \bibnamefont {Chubukov}},\ }\bibfield  {title} {\enquote {\bibinfo {title}
  {Low-energy microscopic models for iron-based superconductors: a review},}\
  }\href {\doibase 10.1088/1361-6633/80/1/014503} {\bibfield  {journal}
  {\bibinfo  {journal} {Reports on Progress in Physics}\ }\textbf {\bibinfo
  {volume} {80}},\ \bibinfo {pages} {014503} (\bibinfo {year}
  {2016})}\BibitemShut {NoStop}%
\bibitem [{\citenamefont {Dai}(2015)}]{RevModPhys.87.855}%
  \BibitemOpen
  \bibfield  {author} {\bibinfo {author} {\bibfnamefont {Pengcheng}\
  \bibnamefont {Dai}},\ }\bibfield  {title} {\enquote {\bibinfo {title}
  {Antiferromagnetic order and spin dynamics in iron-based superconductors},}\
  }\href {\doibase 10.1103/RevModPhys.87.855} {\bibfield  {journal} {\bibinfo
  {journal} {Rev. Mod. Phys.}\ }\textbf {\bibinfo {volume} {87}},\ \bibinfo
  {pages} {855--896} (\bibinfo {year} {2015})}\BibitemShut {NoStop}%
\bibitem [{\citenamefont {Böhmer}\ and\ \citenamefont
  {Kreisel}(2017)}]{Böhmer_2018}%
  \BibitemOpen
  \bibfield  {author} {\bibinfo {author} {\bibfnamefont {Anna~E}\ \bibnamefont
  {Böhmer}}\ and\ \bibinfo {author} {\bibfnamefont {Andreas}\ \bibnamefont
  {Kreisel}},\ }\bibfield  {title} {\enquote {\bibinfo {title} {Nematicity,
  magnetism and superconductivity in {FeSe}},}\ }\href {\doibase
  10.1088/1361-648X/aa9caa} {\bibfield  {journal} {\bibinfo  {journal} {Journal
  of Physics: Condensed Matter}\ }\textbf {\bibinfo {volume} {30}},\ \bibinfo
  {pages} {023001} (\bibinfo {year} {2017})}\BibitemShut {NoStop}%
\bibitem [{\citenamefont {Mazin}\ \emph {et~al.}(2008)\citenamefont {Mazin},
  \citenamefont {Singh}, \citenamefont {Johannes},\ and\ \citenamefont
  {Du}}]{PhysRevLett.101.057003}%
  \BibitemOpen
  \bibfield  {author} {\bibinfo {author} {\bibfnamefont {I.~I.}\ \bibnamefont
  {Mazin}}, \bibinfo {author} {\bibfnamefont {D.~J.}\ \bibnamefont {Singh}},
  \bibinfo {author} {\bibfnamefont {M.~D.}\ \bibnamefont {Johannes}}, \ and\
  \bibinfo {author} {\bibfnamefont {M.~H.}\ \bibnamefont {Du}},\ }\bibfield
  {title} {\enquote {\bibinfo {title} {Unconventional superconductivity with a
  sign reversal in the order parameter of
  {${\mathrm{LaFeAsO}}_{1\ensuremath{-}x}{\mathrm{F}}_{x}$}},}\ }\href
  {\doibase 10.1103/PhysRevLett.101.057003} {\bibfield  {journal} {\bibinfo
  {journal} {Phys. Rev. Lett.}\ }\textbf {\bibinfo {volume} {101}},\ \bibinfo
  {pages} {057003} (\bibinfo {year} {2008})}\BibitemShut {NoStop}%
\bibitem [{\citenamefont {Kuroki}\ \emph {et~al.}(2008)\citenamefont {Kuroki},
  \citenamefont {Onari}, \citenamefont {Arita}, \citenamefont {Usui},
  \citenamefont {Tanaka}, \citenamefont {Kontani},\ and\ \citenamefont
  {Aoki}}]{PhysRevLett.101.087004}%
  \BibitemOpen
  \bibfield  {author} {\bibinfo {author} {\bibfnamefont {Kazuhiko}\
  \bibnamefont {Kuroki}}, \bibinfo {author} {\bibfnamefont {Seiichiro}\
  \bibnamefont {Onari}}, \bibinfo {author} {\bibfnamefont {Ryotaro}\
  \bibnamefont {Arita}}, \bibinfo {author} {\bibfnamefont {Hidetomo}\
  \bibnamefont {Usui}}, \bibinfo {author} {\bibfnamefont {Yukio}\ \bibnamefont
  {Tanaka}}, \bibinfo {author} {\bibfnamefont {Hiroshi}\ \bibnamefont
  {Kontani}}, \ and\ \bibinfo {author} {\bibfnamefont {Hideo}\ \bibnamefont
  {Aoki}},\ }\bibfield  {title} {\enquote {\bibinfo {title} {Unconventional
  pairing originating from the disconnected fermi surfaces of superconducting
  {${\mathrm{LaFeAsO}}_{1\ensuremath{-}x}{\mathrm{F}}_{x}$}},}\ }\href
  {\doibase 10.1103/PhysRevLett.101.087004} {\bibfield  {journal} {\bibinfo
  {journal} {Phys. Rev. Lett.}\ }\textbf {\bibinfo {volume} {101}},\ \bibinfo
  {pages} {087004} (\bibinfo {year} {2008})}\BibitemShut {NoStop}%
\bibitem [{\citenamefont {Seo}\ \emph {et~al.}(2008)\citenamefont {Seo},
  \citenamefont {Bernevig},\ and\ \citenamefont {Hu}}]{PhysRevLett.101.206404}%
  \BibitemOpen
  \bibfield  {author} {\bibinfo {author} {\bibfnamefont {Kangjun}\ \bibnamefont
  {Seo}}, \bibinfo {author} {\bibfnamefont {B.~Andrei}\ \bibnamefont
  {Bernevig}}, \ and\ \bibinfo {author} {\bibfnamefont {Jiangping}\
  \bibnamefont {Hu}},\ }\bibfield  {title} {\enquote {\bibinfo {title} {Pairing
  symmetry in a two-orbital exchange coupling model of oxypnictides},}\ }\href
  {\doibase 10.1103/PhysRevLett.101.206404} {\bibfield  {journal} {\bibinfo
  {journal} {Phys. Rev. Lett.}\ }\textbf {\bibinfo {volume} {101}},\ \bibinfo
  {pages} {206404} (\bibinfo {year} {2008})}\BibitemShut {NoStop}%
\bibitem [{\citenamefont {Fang}\ \emph {et~al.}(2011)\citenamefont {Fang},
  \citenamefont {Wu}, \citenamefont {Thomale}, \citenamefont {Bernevig},\ and\
  \citenamefont {Hu}}]{PhysRevX.1.011009}%
  \BibitemOpen
  \bibfield  {author} {\bibinfo {author} {\bibfnamefont {Chen}\ \bibnamefont
  {Fang}}, \bibinfo {author} {\bibfnamefont {Yang-Le}\ \bibnamefont {Wu}},
  \bibinfo {author} {\bibfnamefont {Ronny}\ \bibnamefont {Thomale}}, \bibinfo
  {author} {\bibfnamefont {B.~Andrei}\ \bibnamefont {Bernevig}}, \ and\
  \bibinfo {author} {\bibfnamefont {Jiangping}\ \bibnamefont {Hu}},\ }\bibfield
   {title} {\enquote {\bibinfo {title} {Robustness of $s$-wave pairing in
  electron-overdoped
  {${A}_{1\ensuremath{-}y}{\mathrm{Fe}}_{2\ensuremath{-}x}{\mathrm{Se}}_{2}$
  $(A\mathbf{=}\mathbf{K},\text{\hskip-0.22em}\mathrm{Cs})$}},}\ }\href
  {\doibase 10.1103/PhysRevX.1.011009} {\bibfield  {journal} {\bibinfo
  {journal} {Phys. Rev. X}\ }\textbf {\bibinfo {volume} {1}},\ \bibinfo {pages}
  {011009} (\bibinfo {year} {2011})}\BibitemShut {NoStop}%
\bibitem [{\citenamefont {Qing-Yan}\ \emph {et~al.}(2012)\citenamefont
  {Qing-Yan}, \citenamefont {Zhi}, \citenamefont {Wen-Hao}, \citenamefont
  {Zuo-Cheng}, \citenamefont {Jin-Song}, \citenamefont {Wei}, \citenamefont
  {Hao}, \citenamefont {Yun-Bo}, \citenamefont {Peng}, \citenamefont {Kai},
  \citenamefont {Jing}, \citenamefont {Can-Li}, \citenamefont {Ke},
  \citenamefont {Jin-Feng}, \citenamefont {Shuai-Hua}, \citenamefont {Ya-Yu},
  \citenamefont {Li-Li}, \citenamefont {Xi}, \citenamefont {Xu-Cun},\ and\
  \citenamefont {Qi-Kun}}]{FeSe/STO}%
  \BibitemOpen
  \bibfield  {author} {\bibinfo {author} {\bibfnamefont {Wang}\ \bibnamefont
  {Qing-Yan}}, \bibinfo {author} {\bibfnamefont {Li}~\bibnamefont {Zhi}},
  \bibinfo {author} {\bibfnamefont {Zhang}\ \bibnamefont {Wen-Hao}}, \bibinfo
  {author} {\bibfnamefont {Zhang}\ \bibnamefont {Zuo-Cheng}}, \bibinfo {author}
  {\bibfnamefont {Zhang}\ \bibnamefont {Jin-Song}}, \bibinfo {author}
  {\bibfnamefont {Li}~\bibnamefont {Wei}}, \bibinfo {author} {\bibfnamefont
  {Ding}\ \bibnamefont {Hao}}, \bibinfo {author} {\bibfnamefont
  {Ou}~\bibnamefont {Yun-Bo}}, \bibinfo {author} {\bibfnamefont {Deng}\
  \bibnamefont {Peng}}, \bibinfo {author} {\bibfnamefont {Chang}\ \bibnamefont
  {Kai}}, \bibinfo {author} {\bibfnamefont {Wen}\ \bibnamefont {Jing}},
  \bibinfo {author} {\bibfnamefont {Song}\ \bibnamefont {Can-Li}}, \bibinfo
  {author} {\bibfnamefont {He}~\bibnamefont {Ke}}, \bibinfo {author}
  {\bibfnamefont {Jia}\ \bibnamefont {Jin-Feng}}, \bibinfo {author}
  {\bibfnamefont {Ji}~\bibnamefont {Shuai-Hua}}, \bibinfo {author}
  {\bibfnamefont {Wang}\ \bibnamefont {Ya-Yu}}, \bibinfo {author}
  {\bibfnamefont {Wang}\ \bibnamefont {Li-Li}}, \bibinfo {author}
  {\bibfnamefont {Chen}\ \bibnamefont {Xi}}, \bibinfo {author} {\bibfnamefont
  {Ma}~\bibnamefont {Xu-Cun}}, \ and\ \bibinfo {author} {\bibfnamefont {Xue}\
  \bibnamefont {Qi-Kun}},\ }\bibfield  {title} {\enquote {\bibinfo {title}
  {Interface-induced high-temperature superconductivity in single unit-cell
  {FeSe} films on {SrTiO$_3$}},}\ }\href {\doibase
  10.1088/0256-307X/29/3/037402} {\bibfield  {journal} {\bibinfo  {journal}
  {Chin. Phys. Lett.}\ }\textbf {\bibinfo {volume} {29}},\ \bibinfo {pages}
  {037402--037402} (\bibinfo {year} {2012})}\BibitemShut {NoStop}%
\bibitem [{\citenamefont {He}\ \emph {et~al.}(2013)\citenamefont {He},
  \citenamefont {He}, \citenamefont {Zhang}, \citenamefont {Zhao},
  \citenamefont {Liu}, \citenamefont {Liu}, \citenamefont {Mou}, \citenamefont
  {Ou}, \citenamefont {Wang}, \citenamefont {Li}, \citenamefont {Wang},
  \citenamefont {Peng}, \citenamefont {Liu}, \citenamefont {Chen},
  \citenamefont {Yu}, \citenamefont {Liu}, \citenamefont {Dong}, \citenamefont
  {Zhang}, \citenamefont {Chen}, \citenamefont {Xu}, \citenamefont {Chen},
  \citenamefont {Ma}, \citenamefont {Xue},\ and\ \citenamefont
  {Zhou}}]{He2013}%
  \BibitemOpen
  \bibfield  {author} {\bibinfo {author} {\bibfnamefont {Shaolong}\
  \bibnamefont {He}}, \bibinfo {author} {\bibfnamefont {Junfeng}\ \bibnamefont
  {He}}, \bibinfo {author} {\bibfnamefont {Wenhao}\ \bibnamefont {Zhang}},
  \bibinfo {author} {\bibfnamefont {Lin}\ \bibnamefont {Zhao}}, \bibinfo
  {author} {\bibfnamefont {Defa}\ \bibnamefont {Liu}}, \bibinfo {author}
  {\bibfnamefont {Xu}~\bibnamefont {Liu}}, \bibinfo {author} {\bibfnamefont
  {Daixiang}\ \bibnamefont {Mou}}, \bibinfo {author} {\bibfnamefont {Yun-Bo}\
  \bibnamefont {Ou}}, \bibinfo {author} {\bibfnamefont {Qing-Yan}\ \bibnamefont
  {Wang}}, \bibinfo {author} {\bibfnamefont {Zhi}\ \bibnamefont {Li}}, \bibinfo
  {author} {\bibfnamefont {Lili}\ \bibnamefont {Wang}}, \bibinfo {author}
  {\bibfnamefont {Yingying}\ \bibnamefont {Peng}}, \bibinfo {author}
  {\bibfnamefont {Yan}\ \bibnamefont {Liu}}, \bibinfo {author} {\bibfnamefont
  {Chaoyu}\ \bibnamefont {Chen}}, \bibinfo {author} {\bibfnamefont
  {Li}~\bibnamefont {Yu}}, \bibinfo {author} {\bibfnamefont {Guodong}\
  \bibnamefont {Liu}}, \bibinfo {author} {\bibfnamefont {Xiaoli}\ \bibnamefont
  {Dong}}, \bibinfo {author} {\bibfnamefont {Jun}\ \bibnamefont {Zhang}},
  \bibinfo {author} {\bibfnamefont {Chuangtian}\ \bibnamefont {Chen}}, \bibinfo
  {author} {\bibfnamefont {Zuyan}\ \bibnamefont {Xu}}, \bibinfo {author}
  {\bibfnamefont {Xi}~\bibnamefont {Chen}}, \bibinfo {author} {\bibfnamefont
  {Xucun}\ \bibnamefont {Ma}}, \bibinfo {author} {\bibfnamefont {Qikun}\
  \bibnamefont {Xue}}, \ and\ \bibinfo {author} {\bibfnamefont {X.J.}\
  \bibnamefont {Zhou}},\ }\bibfield  {title} {\enquote {\bibinfo {title} {Phase
  diagram and electronic indication of high-temperature superconductivity at 65
  {K} in single-layer {FeSe} films},}\ }\href {\doibase 10.1038/nmat3648}
  {\bibfield  {journal} {\bibinfo  {journal} {Nature Materials}\ }\textbf
  {\bibinfo {volume} {12}},\ \bibinfo {pages} {605--610} (\bibinfo {year}
  {2013})}\BibitemShut {NoStop}%
\bibitem [{\citenamefont {Tan}\ \emph {et~al.}(2013)\citenamefont {Tan},
  \citenamefont {Zhang}, \citenamefont {Xia}, \citenamefont {Ye}, \citenamefont
  {Chen}, \citenamefont {Xie}, \citenamefont {Peng}, \citenamefont {Xu},
  \citenamefont {Fan}, \citenamefont {Xu}, \citenamefont {Jiang}, \citenamefont
  {Zhang}, \citenamefont {Lai}, \citenamefont {Xiang}, \citenamefont {Hu},
  \citenamefont {Xie},\ and\ \citenamefont {Feng}}]{Tan2013}%
  \BibitemOpen
  \bibfield  {author} {\bibinfo {author} {\bibfnamefont {Shiyong}\ \bibnamefont
  {Tan}}, \bibinfo {author} {\bibfnamefont {Yan}\ \bibnamefont {Zhang}},
  \bibinfo {author} {\bibfnamefont {Miao}\ \bibnamefont {Xia}}, \bibinfo
  {author} {\bibfnamefont {Zirong}\ \bibnamefont {Ye}}, \bibinfo {author}
  {\bibfnamefont {Fei}\ \bibnamefont {Chen}}, \bibinfo {author} {\bibfnamefont
  {Xin}\ \bibnamefont {Xie}}, \bibinfo {author} {\bibfnamefont {Rui}\
  \bibnamefont {Peng}}, \bibinfo {author} {\bibfnamefont {Difei}\ \bibnamefont
  {Xu}}, \bibinfo {author} {\bibfnamefont {Qin}\ \bibnamefont {Fan}}, \bibinfo
  {author} {\bibfnamefont {Haichao}\ \bibnamefont {Xu}}, \bibinfo {author}
  {\bibfnamefont {Juan}\ \bibnamefont {Jiang}}, \bibinfo {author}
  {\bibfnamefont {Tong}\ \bibnamefont {Zhang}}, \bibinfo {author}
  {\bibfnamefont {Xinchun}\ \bibnamefont {Lai}}, \bibinfo {author}
  {\bibfnamefont {Tao}\ \bibnamefont {Xiang}}, \bibinfo {author} {\bibfnamefont
  {Jiangping}\ \bibnamefont {Hu}}, \bibinfo {author} {\bibfnamefont {Binping}\
  \bibnamefont {Xie}}, \ and\ \bibinfo {author} {\bibfnamefont {Donglai}\
  \bibnamefont {Feng}},\ }\bibfield  {title} {\enquote {\bibinfo {title}
  {Interface-induced superconductivity and strain-dependent spin density waves
  in {FeSe/SrTiO$_3$} thin films},}\ }\href {\doibase 10.1038/nmat3654}
  {\bibfield  {journal} {\bibinfo  {journal} {Nature Materials}\ }\textbf
  {\bibinfo {volume} {12}},\ \bibinfo {pages} {634--640} (\bibinfo {year}
  {2013})}\BibitemShut {NoStop}%
\bibitem [{\citenamefont {Lee}\ \emph {et~al.}(2014)\citenamefont {Lee},
  \citenamefont {Schmitt}, \citenamefont {Moore}, \citenamefont {Johnston},
  \citenamefont {Cui}, \citenamefont {Li}, \citenamefont {Yi}, \citenamefont
  {Liu}, \citenamefont {Hashimoto}, \citenamefont {Zhang}, \citenamefont {Lu},
  \citenamefont {Devereaux}, \citenamefont {Lee},\ and\ \citenamefont
  {Shen}}]{Lee2014}%
  \BibitemOpen
  \bibfield  {author} {\bibinfo {author} {\bibfnamefont {J.~J.}\ \bibnamefont
  {Lee}}, \bibinfo {author} {\bibfnamefont {F.~T.}\ \bibnamefont {Schmitt}},
  \bibinfo {author} {\bibfnamefont {R.~G.}\ \bibnamefont {Moore}}, \bibinfo
  {author} {\bibfnamefont {S.}~\bibnamefont {Johnston}}, \bibinfo {author}
  {\bibfnamefont {Y.-T.}\ \bibnamefont {Cui}}, \bibinfo {author} {\bibfnamefont
  {W.}~\bibnamefont {Li}}, \bibinfo {author} {\bibfnamefont {M.}~\bibnamefont
  {Yi}}, \bibinfo {author} {\bibfnamefont {Z.~K.}\ \bibnamefont {Liu}},
  \bibinfo {author} {\bibfnamefont {M.}~\bibnamefont {Hashimoto}}, \bibinfo
  {author} {\bibfnamefont {Y.}~\bibnamefont {Zhang}}, \bibinfo {author}
  {\bibfnamefont {D.~H.}\ \bibnamefont {Lu}}, \bibinfo {author} {\bibfnamefont
  {T.~P.}\ \bibnamefont {Devereaux}}, \bibinfo {author} {\bibfnamefont {D.-H.}\
  \bibnamefont {Lee}}, \ and\ \bibinfo {author} {\bibfnamefont {Z.-X.}\
  \bibnamefont {Shen}},\ }\bibfield  {title} {\enquote {\bibinfo {title}
  {Interfacial mode coupling as the origin of the enhancement of {$T_c$} in
  {FeSe} films on {SrTiO$_3$}},}\ }\href {\doibase 10.1038/nature13894}
  {\bibfield  {journal} {\bibinfo  {journal} {Nature}\ }\textbf {\bibinfo
  {volume} {515}},\ \bibinfo {pages} {245--248} (\bibinfo {year}
  {2014})}\BibitemShut {NoStop}%
\bibitem [{\citenamefont {Zhang}\ \emph {et~al.}(2015)\citenamefont {Zhang},
  \citenamefont {Wang}, \citenamefont {Song}, \citenamefont {Liu},
  \citenamefont {Peng}, \citenamefont {Moler}, \citenamefont {Feng},\ and\
  \citenamefont {Wang}}]{ZHANG20151301}%
  \BibitemOpen
  \bibfield  {author} {\bibinfo {author} {\bibfnamefont {Zuocheng}\
  \bibnamefont {Zhang}}, \bibinfo {author} {\bibfnamefont {Yi-Hua}\
  \bibnamefont {Wang}}, \bibinfo {author} {\bibfnamefont {Qi}~\bibnamefont
  {Song}}, \bibinfo {author} {\bibfnamefont {Chang}\ \bibnamefont {Liu}},
  \bibinfo {author} {\bibfnamefont {Rui}\ \bibnamefont {Peng}}, \bibinfo
  {author} {\bibfnamefont {K.A.}\ \bibnamefont {Moler}}, \bibinfo {author}
  {\bibfnamefont {Donglai}\ \bibnamefont {Feng}}, \ and\ \bibinfo {author}
  {\bibfnamefont {Yayu}\ \bibnamefont {Wang}},\ }\bibfield  {title} {\enquote
  {\bibinfo {title} {Onset of the meissner effect at 65{K} in {FeSe} thin film
  grown on {Nb}-doped {SrTiO$_3$} substrate},}\ }\href {\doibase
  https://doi.org/10.1007/s11434-015-0842-8} {\bibfield  {journal} {\bibinfo
  {journal} {Science Bulletin}\ }\textbf {\bibinfo {volume} {60}},\ \bibinfo
  {pages} {1301--1304} (\bibinfo {year} {2015})}\BibitemShut {NoStop}%
\bibitem [{\citenamefont {Ge}\ \emph {et~al.}(2015)\citenamefont {Ge},
  \citenamefont {Liu}, \citenamefont {Liu}, \citenamefont {Gao}, \citenamefont
  {Qian}, \citenamefont {Xue}, \citenamefont {Liu},\ and\ \citenamefont
  {Jia}}]{Ge2015}%
  \BibitemOpen
  \bibfield  {author} {\bibinfo {author} {\bibfnamefont {Jian-Feng}\
  \bibnamefont {Ge}}, \bibinfo {author} {\bibfnamefont {Zhi-Long}\ \bibnamefont
  {Liu}}, \bibinfo {author} {\bibfnamefont {Canhua}\ \bibnamefont {Liu}},
  \bibinfo {author} {\bibfnamefont {Chun-Lei}\ \bibnamefont {Gao}}, \bibinfo
  {author} {\bibfnamefont {Dong}\ \bibnamefont {Qian}}, \bibinfo {author}
  {\bibfnamefont {Qi-Kun}\ \bibnamefont {Xue}}, \bibinfo {author}
  {\bibfnamefont {Ying}\ \bibnamefont {Liu}}, \ and\ \bibinfo {author}
  {\bibfnamefont {Jin-Feng}\ \bibnamefont {Jia}},\ }\bibfield  {title}
  {\enquote {\bibinfo {title} {Superconductivity above 100 {K} in single-layer
  {FeSe} films on doped {SrTiO$_3$}},}\ }\href {\doibase 10.1038/nmat4153}
  {\bibfield  {journal} {\bibinfo  {journal} {Nature Materials}\ }\textbf
  {\bibinfo {volume} {14}},\ \bibinfo {pages} {285--289} (\bibinfo {year}
  {2015})}\BibitemShut {NoStop}%
\bibitem [{\citenamefont {Huang}\ and\ \citenamefont
  {Hoffman}(2017)}]{annurev:/content/journals/10.1146/annurev-conmatphys-031016-025242}%
  \BibitemOpen
  \bibfield  {author} {\bibinfo {author} {\bibfnamefont {Dennis}\ \bibnamefont
  {Huang}}\ and\ \bibinfo {author} {\bibfnamefont {Jennifer~E.}\ \bibnamefont
  {Hoffman}},\ }\bibfield  {title} {\enquote {\bibinfo {title} {Monolayer
  {FeSe} on {SrTiO$_3$}},}\ }\href {\doibase
  https://doi.org/10.1146/annurev-conmatphys-031016-025242} {\bibfield
  {journal} {\bibinfo  {journal} {Annual Review of Condensed Matter Physics}\
  }\textbf {\bibinfo {volume} {8}},\ \bibinfo {pages} {311--336} (\bibinfo
  {year} {2017})}\BibitemShut {NoStop}%
\bibitem [{\citenamefont {Xu}\ \emph {et~al.}(2021)\citenamefont {Xu},
  \citenamefont {Rong}, \citenamefont {Wang}, \citenamefont {Wu}, \citenamefont
  {Hu}, \citenamefont {Cai}, \citenamefont {Gao}, \citenamefont {Yan},
  \citenamefont {Li}, \citenamefont {Yin}, \citenamefont {Chen}, \citenamefont
  {Huang}, \citenamefont {Zhu}, \citenamefont {Huang}, \citenamefont {Liu},
  \citenamefont {Xu}, \citenamefont {Zhao},\ and\ \citenamefont
  {Zhou}}]{Xu2021}%
  \BibitemOpen
  \bibfield  {author} {\bibinfo {author} {\bibfnamefont {Yu}~\bibnamefont
  {Xu}}, \bibinfo {author} {\bibfnamefont {Hongtao}\ \bibnamefont {Rong}},
  \bibinfo {author} {\bibfnamefont {Qingyan}\ \bibnamefont {Wang}}, \bibinfo
  {author} {\bibfnamefont {Dingsong}\ \bibnamefont {Wu}}, \bibinfo {author}
  {\bibfnamefont {Yong}\ \bibnamefont {Hu}}, \bibinfo {author} {\bibfnamefont
  {Yongqing}\ \bibnamefont {Cai}}, \bibinfo {author} {\bibfnamefont {Qiang}\
  \bibnamefont {Gao}}, \bibinfo {author} {\bibfnamefont {Hongtao}\ \bibnamefont
  {Yan}}, \bibinfo {author} {\bibfnamefont {Cong}\ \bibnamefont {Li}}, \bibinfo
  {author} {\bibfnamefont {Chaohui}\ \bibnamefont {Yin}}, \bibinfo {author}
  {\bibfnamefont {Hao}\ \bibnamefont {Chen}}, \bibinfo {author} {\bibfnamefont
  {Jianwei}\ \bibnamefont {Huang}}, \bibinfo {author} {\bibfnamefont {Zhihai}\
  \bibnamefont {Zhu}}, \bibinfo {author} {\bibfnamefont {Yuan}\ \bibnamefont
  {Huang}}, \bibinfo {author} {\bibfnamefont {Guodong}\ \bibnamefont {Liu}},
  \bibinfo {author} {\bibfnamefont {Zuyan}\ \bibnamefont {Xu}}, \bibinfo
  {author} {\bibfnamefont {Lin}\ \bibnamefont {Zhao}}, \ and\ \bibinfo {author}
  {\bibfnamefont {X.~J.}\ \bibnamefont {Zhou}},\ }\bibfield  {title} {\enquote
  {\bibinfo {title} {Spectroscopic evidence of superconductivity pairing at 83
  {K} in single-layer {FeSe/SrTiO$_3$} films},}\ }\href {\doibase
  10.1038/s41467-021-23106-y} {\bibfield  {journal} {\bibinfo  {journal}
  {Nature Communications}\ }\textbf {\bibinfo {volume} {12}},\ \bibinfo {pages}
  {2840} (\bibinfo {year} {2021})}\BibitemShut {NoStop}%
\bibitem [{\citenamefont {Liu}\ \emph {et~al.}(2012)\citenamefont {Liu},
  \citenamefont {Zhang}, \citenamefont {Mou}, \citenamefont {He}, \citenamefont
  {Ou}, \citenamefont {Wang}, \citenamefont {Li}, \citenamefont {Wang},
  \citenamefont {Zhao}, \citenamefont {He}, \citenamefont {Peng}, \citenamefont
  {Liu}, \citenamefont {Chen}, \citenamefont {Yu}, \citenamefont {Liu},
  \citenamefont {Dong}, \citenamefont {Zhang}, \citenamefont {Chen},
  \citenamefont {Xu}, \citenamefont {Hu}, \citenamefont {Chen}, \citenamefont
  {Ma}, \citenamefont {Xue},\ and\ \citenamefont {Zhou}}]{Liu2012FeSeARPES}%
  \BibitemOpen
  \bibfield  {author} {\bibinfo {author} {\bibfnamefont {Defa}\ \bibnamefont
  {Liu}}, \bibinfo {author} {\bibfnamefont {Wenhao}\ \bibnamefont {Zhang}},
  \bibinfo {author} {\bibfnamefont {Daixiang}\ \bibnamefont {Mou}}, \bibinfo
  {author} {\bibfnamefont {Junfeng}\ \bibnamefont {He}}, \bibinfo {author}
  {\bibfnamefont {Yun-Bo}\ \bibnamefont {Ou}}, \bibinfo {author} {\bibfnamefont
  {Qing-Yan}\ \bibnamefont {Wang}}, \bibinfo {author} {\bibfnamefont {Zhi}\
  \bibnamefont {Li}}, \bibinfo {author} {\bibfnamefont {Lili}\ \bibnamefont
  {Wang}}, \bibinfo {author} {\bibfnamefont {Lin}\ \bibnamefont {Zhao}},
  \bibinfo {author} {\bibfnamefont {Shaolong}\ \bibnamefont {He}}, \bibinfo
  {author} {\bibfnamefont {Yingying}\ \bibnamefont {Peng}}, \bibinfo {author}
  {\bibfnamefont {Xu}~\bibnamefont {Liu}}, \bibinfo {author} {\bibfnamefont
  {Chaoyu}\ \bibnamefont {Chen}}, \bibinfo {author} {\bibfnamefont
  {Li}~\bibnamefont {Yu}}, \bibinfo {author} {\bibfnamefont {Guodong}\
  \bibnamefont {Liu}}, \bibinfo {author} {\bibfnamefont {Xiaoli}\ \bibnamefont
  {Dong}}, \bibinfo {author} {\bibfnamefont {Jun}\ \bibnamefont {Zhang}},
  \bibinfo {author} {\bibfnamefont {Chuangtian}\ \bibnamefont {Chen}}, \bibinfo
  {author} {\bibfnamefont {Zuyan}\ \bibnamefont {Xu}}, \bibinfo {author}
  {\bibfnamefont {Jiangping}\ \bibnamefont {Hu}}, \bibinfo {author}
  {\bibfnamefont {Xi}~\bibnamefont {Chen}}, \bibinfo {author} {\bibfnamefont
  {Xucun}\ \bibnamefont {Ma}}, \bibinfo {author} {\bibfnamefont {Qikun}\
  \bibnamefont {Xue}}, \ and\ \bibinfo {author} {\bibfnamefont {X.J.}\
  \bibnamefont {Zhou}},\ }\bibfield  {title} {\enquote {\bibinfo {title}
  {Electronic origin of high-temperature superconductivity in single-layer
  $\mathrm{FeSe}$ superconductor},}\ }\href {\doibase 10.1038/ncomms1946}
  {\bibfield  {journal} {\bibinfo  {journal} {Nature Communications}\ }\textbf
  {\bibinfo {volume} {3}},\ \bibinfo {pages} {931} (\bibinfo {year}
  {2012})}\BibitemShut {NoStop}%
\bibitem [{\citenamefont {Fan}\ \emph {et~al.}(2015)\citenamefont {Fan},
  \citenamefont {Zhang}, \citenamefont {Liu}, \citenamefont {Yan},
  \citenamefont {Ren}, \citenamefont {Peng}, \citenamefont {Xu}, \citenamefont
  {Xie}, \citenamefont {Hu}, \citenamefont {Zhang},\ and\ \citenamefont
  {Feng}}]{Fan2015}%
  \BibitemOpen
  \bibfield  {author} {\bibinfo {author} {\bibfnamefont {Q.}~\bibnamefont
  {Fan}}, \bibinfo {author} {\bibfnamefont {W.~H.}\ \bibnamefont {Zhang}},
  \bibinfo {author} {\bibfnamefont {X.}~\bibnamefont {Liu}}, \bibinfo {author}
  {\bibfnamefont {Y.~J.}\ \bibnamefont {Yan}}, \bibinfo {author} {\bibfnamefont
  {M.~Q.}\ \bibnamefont {Ren}}, \bibinfo {author} {\bibfnamefont
  {R.}~\bibnamefont {Peng}}, \bibinfo {author} {\bibfnamefont {H.~C.}\
  \bibnamefont {Xu}}, \bibinfo {author} {\bibfnamefont {B.~P.}\ \bibnamefont
  {Xie}}, \bibinfo {author} {\bibfnamefont {J.~P.}\ \bibnamefont {Hu}},
  \bibinfo {author} {\bibfnamefont {T.}~\bibnamefont {Zhang}}, \ and\ \bibinfo
  {author} {\bibfnamefont {D.~L.}\ \bibnamefont {Feng}},\ }\bibfield  {title}
  {\enquote {\bibinfo {title} {Plain $s$-wave superconductivity in single-layer
  {FeSe} on {SrTiO$_3$} probed by scanning tunnelling microscopy},}\ }\href
  {\doibase 10.1038/nphys3450} {\bibfield  {journal} {\bibinfo  {journal}
  {Nature Physics}\ }\textbf {\bibinfo {volume} {11}},\ \bibinfo {pages}
  {946--952} (\bibinfo {year} {2015})}\BibitemShut {NoStop}%
\bibitem [{\citenamefont {Chen}\ \emph {et~al.}(2015)\citenamefont {Chen},
  \citenamefont {Maiti}, \citenamefont {Linscheid},\ and\ \citenamefont
  {Hirschfeld}}]{PhysRevB.92.224514}%
  \BibitemOpen
  \bibfield  {author} {\bibinfo {author} {\bibfnamefont {Xiao}\ \bibnamefont
  {Chen}}, \bibinfo {author} {\bibfnamefont {S.}~\bibnamefont {Maiti}},
  \bibinfo {author} {\bibfnamefont {A.}~\bibnamefont {Linscheid}}, \ and\
  \bibinfo {author} {\bibfnamefont {P.~J.}\ \bibnamefont {Hirschfeld}},\
  }\bibfield  {title} {\enquote {\bibinfo {title} {Electron pairing in the
  presence of incipient bands in iron-based superconductors},}\ }\href
  {\doibase 10.1103/PhysRevB.92.224514} {\bibfield  {journal} {\bibinfo
  {journal} {Phys. Rev. B}\ }\textbf {\bibinfo {volume} {92}},\ \bibinfo
  {pages} {224514} (\bibinfo {year} {2015})}\BibitemShut {NoStop}%
\bibitem [{\citenamefont {Huang}\ \emph {et~al.}(2016)\citenamefont {Huang},
  \citenamefont {Webb}, \citenamefont {Fang}, \citenamefont {Song},
  \citenamefont {Chang}, \citenamefont {Moodera}, \citenamefont {Kaxiras},\
  and\ \citenamefont {Hoffman}}]{PhysRevB.93.125129}%
  \BibitemOpen
  \bibfield  {author} {\bibinfo {author} {\bibfnamefont {Dennis}\ \bibnamefont
  {Huang}}, \bibinfo {author} {\bibfnamefont {Tatiana~A.}\ \bibnamefont
  {Webb}}, \bibinfo {author} {\bibfnamefont {Shiang}\ \bibnamefont {Fang}},
  \bibinfo {author} {\bibfnamefont {Can-Li}\ \bibnamefont {Song}}, \bibinfo
  {author} {\bibfnamefont {Cui-Zu}\ \bibnamefont {Chang}}, \bibinfo {author}
  {\bibfnamefont {Jagadeesh~S.}\ \bibnamefont {Moodera}}, \bibinfo {author}
  {\bibfnamefont {Efthimios}\ \bibnamefont {Kaxiras}}, \ and\ \bibinfo {author}
  {\bibfnamefont {Jennifer~E.}\ \bibnamefont {Hoffman}},\ }\bibfield  {title}
  {\enquote {\bibinfo {title} {Bounds on nanoscale nematicity in single-layer
  {$\mathrm{FeSe}/{\mathrm{SrTiO}}_{3}$}},}\ }\href {\doibase
  10.1103/PhysRevB.93.125129} {\bibfield  {journal} {\bibinfo  {journal} {Phys.
  Rev. B}\ }\textbf {\bibinfo {volume} {93}},\ \bibinfo {pages} {125129}
  (\bibinfo {year} {2016})}\BibitemShut {NoStop}%
\bibitem [{\citenamefont {Linscheid}\ \emph {et~al.}(2016)\citenamefont
  {Linscheid}, \citenamefont {Maiti}, \citenamefont {Wang}, \citenamefont
  {Johnston},\ and\ \citenamefont {Hirschfeld}}]{PhysRevLett.117.077003}%
  \BibitemOpen
  \bibfield  {author} {\bibinfo {author} {\bibfnamefont {A.}~\bibnamefont
  {Linscheid}}, \bibinfo {author} {\bibfnamefont {S.}~\bibnamefont {Maiti}},
  \bibinfo {author} {\bibfnamefont {Y.}~\bibnamefont {Wang}}, \bibinfo {author}
  {\bibfnamefont {S.}~\bibnamefont {Johnston}}, \ and\ \bibinfo {author}
  {\bibfnamefont {P.~J.}\ \bibnamefont {Hirschfeld}},\ }\bibfield  {title}
  {\enquote {\bibinfo {title} {High {${T}_{c}$} via spin fluctuations from
  incipient bands: Application to monolayers and intercalates of {FeSe}},}\
  }\href {\doibase 10.1103/PhysRevLett.117.077003} {\bibfield  {journal}
  {\bibinfo  {journal} {Phys. Rev. Lett.}\ }\textbf {\bibinfo {volume} {117}},\
  \bibinfo {pages} {077003} (\bibinfo {year} {2016})}\BibitemShut {NoStop}%
\bibitem [{\citenamefont {Zhang}\ \emph {et~al.}(2016)\citenamefont {Zhang},
  \citenamefont {Lee}, \citenamefont {Moore}, \citenamefont {Li}, \citenamefont
  {Yi}, \citenamefont {Hashimoto}, \citenamefont {Lu}, \citenamefont
  {Devereaux}, \citenamefont {Lee},\ and\ \citenamefont
  {Shen}}]{PhysRevLett.117.117001}%
  \BibitemOpen
  \bibfield  {author} {\bibinfo {author} {\bibfnamefont {Y.}~\bibnamefont
  {Zhang}}, \bibinfo {author} {\bibfnamefont {J.~J.}\ \bibnamefont {Lee}},
  \bibinfo {author} {\bibfnamefont {R.~G.}\ \bibnamefont {Moore}}, \bibinfo
  {author} {\bibfnamefont {W.}~\bibnamefont {Li}}, \bibinfo {author}
  {\bibfnamefont {M.}~\bibnamefont {Yi}}, \bibinfo {author} {\bibfnamefont
  {M.}~\bibnamefont {Hashimoto}}, \bibinfo {author} {\bibfnamefont {D.~H.}\
  \bibnamefont {Lu}}, \bibinfo {author} {\bibfnamefont {T.~P.}\ \bibnamefont
  {Devereaux}}, \bibinfo {author} {\bibfnamefont {D.-H.}\ \bibnamefont {Lee}},
  \ and\ \bibinfo {author} {\bibfnamefont {Z.-X.}\ \bibnamefont {Shen}},\
  }\bibfield  {title} {\enquote {\bibinfo {title} {Superconducting gap
  anisotropy in monolayer {FeSe} thin film},}\ }\href {\doibase
  10.1103/PhysRevLett.117.117001} {\bibfield  {journal} {\bibinfo  {journal}
  {Phys. Rev. Lett.}\ }\textbf {\bibinfo {volume} {117}},\ \bibinfo {pages}
  {117001} (\bibinfo {year} {2016})}\BibitemShut {NoStop}%
\bibitem [{\citenamefont {Ding}\ \emph {et~al.}(2024)\citenamefont {Ding},
  \citenamefont {Xu}, \citenamefont {Jiao}, \citenamefont {Hu}, \citenamefont
  {Zhao}, \citenamefont {Yang}, \citenamefont {Jiang}, \citenamefont {Jia},
  \citenamefont {Wang}, \citenamefont {Hu} \emph
  {et~al.}}]{ding2024sublattice}%
  \BibitemOpen
  \bibfield  {author} {\bibinfo {author} {\bibfnamefont {Cui}\ \bibnamefont
  {Ding}}, \bibinfo {author} {\bibfnamefont {Zhipeng}\ \bibnamefont {Xu}},
  \bibinfo {author} {\bibfnamefont {Xiaotong}\ \bibnamefont {Jiao}}, \bibinfo
  {author} {\bibfnamefont {Qiyin}\ \bibnamefont {Hu}}, \bibinfo {author}
  {\bibfnamefont {Wenxuan}\ \bibnamefont {Zhao}}, \bibinfo {author}
  {\bibfnamefont {Lexian}\ \bibnamefont {Yang}}, \bibinfo {author}
  {\bibfnamefont {Kun}\ \bibnamefont {Jiang}}, \bibinfo {author} {\bibfnamefont
  {Jin-Feng}\ \bibnamefont {Jia}}, \bibinfo {author} {\bibfnamefont {Lili}\
  \bibnamefont {Wang}}, \bibinfo {author} {\bibfnamefont {Jiangping}\
  \bibnamefont {Hu}},  \emph {et~al.},\ }\bibfield  {title} {\enquote {\bibinfo
  {title} {Sublattice dichotomy in monolayer $\mathrm{FeSe}$ superconductor},}\
  }\href {https://arxiv.org/abs/2406.15239} {\bibfield  {journal} {\bibinfo
  {journal} {arXiv:2406.15239}\ } (\bibinfo {year} {2024})}\BibitemShut
  {NoStop}%
\bibitem [{\citenamefont {Cvetkovic}\ and\ \citenamefont
  {Vafek}(2013)}]{vafek_PhysRevB.88.134510}%
  \BibitemOpen
  \bibfield  {author} {\bibinfo {author} {\bibfnamefont {Vladimir}\
  \bibnamefont {Cvetkovic}}\ and\ \bibinfo {author} {\bibfnamefont {Oskar}\
  \bibnamefont {Vafek}},\ }\bibfield  {title} {\enquote {\bibinfo {title}
  {Space group symmetry, spin-orbit coupling, and the low-energy effective
  hamiltonian for iron-based superconductors},}\ }\href {\doibase
  10.1103/PhysRevB.88.134510} {\bibfield  {journal} {\bibinfo  {journal} {Phys.
  Rev. B}\ }\textbf {\bibinfo {volume} {88}},\ \bibinfo {pages} {134510}
  (\bibinfo {year} {2013})}\BibitemShut {NoStop}%
\bibitem [{\citenamefont {Li}\ \emph {et~al.}(2024)\citenamefont {Li},
  \citenamefont {Jiang},\ and\ \citenamefont {Hu}}]{PhysRevB.110.094517}%
  \BibitemOpen
  \bibfield  {author} {\bibinfo {author} {\bibfnamefont {Pengfei}\ \bibnamefont
  {Li}}, \bibinfo {author} {\bibfnamefont {Kun}\ \bibnamefont {Jiang}}, \ and\
  \bibinfo {author} {\bibfnamefont {Jiangping}\ \bibnamefont {Hu}},\ }\bibfield
   {title} {\enquote {\bibinfo {title} {Paramagnetic contribution in
  superconductors with different-mass cooper pairs},}\ }\href {\doibase
  10.1103/PhysRevB.110.094517} {\bibfield  {journal} {\bibinfo  {journal}
  {Phys. Rev. B}\ }\textbf {\bibinfo {volume} {110}},\ \bibinfo {pages}
  {094517} (\bibinfo {year} {2024})}\BibitemShut {NoStop}%
\bibitem [{\citenamefont {Mei}\ \emph {et~al.}(2025)\citenamefont {Mei},
  \citenamefont {Qin},\ and\ \citenamefont {Hu}}]{mei2025interband}%
  \BibitemOpen
  \bibfield  {author} {\bibinfo {author} {\bibfnamefont {Jiong}\ \bibnamefont
  {Mei}}, \bibinfo {author} {\bibfnamefont {Shengshan}\ \bibnamefont {Qin}}, \
  and\ \bibinfo {author} {\bibfnamefont {Jiangping}\ \bibnamefont {Hu}},\
  }\bibfield  {title} {\enquote {\bibinfo {title} {Interband-pairing-boosted
  supercurrent diode effect in multiband superconductors},}\ }\href
  {https://arxiv.org/abs/2510.15788} {\bibfield  {journal} {\bibinfo  {journal}
  {arXiv:2510.15788}\ } (\bibinfo {year} {2025})}\BibitemShut {NoStop}%
\bibitem [{\citenamefont {Moreo}\ \emph {et~al.}(2009)\citenamefont {Moreo},
  \citenamefont {Daghofer}, \citenamefont {Nicholson},\ and\ \citenamefont
  {Dagotto}}]{PhysRevB.80.104507}%
  \BibitemOpen
  \bibfield  {author} {\bibinfo {author} {\bibfnamefont {Adriana}\ \bibnamefont
  {Moreo}}, \bibinfo {author} {\bibfnamefont {Maria}\ \bibnamefont {Daghofer}},
  \bibinfo {author} {\bibfnamefont {Andrew}\ \bibnamefont {Nicholson}}, \ and\
  \bibinfo {author} {\bibfnamefont {Elbio}\ \bibnamefont {Dagotto}},\
  }\bibfield  {title} {\enquote {\bibinfo {title} {Interband pairing in
  multiorbital systems},}\ }\href {\doibase 10.1103/PhysRevB.80.104507}
  {\bibfield  {journal} {\bibinfo  {journal} {Phys. Rev. B}\ }\textbf {\bibinfo
  {volume} {80}},\ \bibinfo {pages} {104507} (\bibinfo {year}
  {2009})}\BibitemShut {NoStop}%
\bibitem [{\citenamefont {Tahir-Kheli}(1998)}]{PhysRevB.58.12307}%
  \BibitemOpen
  \bibfield  {author} {\bibinfo {author} {\bibfnamefont {Jamil}\ \bibnamefont
  {Tahir-Kheli}},\ }\bibfield  {title} {\enquote {\bibinfo {title} {Interband
  pairing theory of superconductivity},}\ }\href {\doibase
  10.1103/PhysRevB.58.12307} {\bibfield  {journal} {\bibinfo  {journal} {Phys.
  Rev. B}\ }\textbf {\bibinfo {volume} {58}},\ \bibinfo {pages} {12307--12322}
  (\bibinfo {year} {1998})}\BibitemShut {NoStop}%
\bibitem [{\citenamefont {Dolgov}\ \emph {et~al.}(1987)\citenamefont {Dolgov},
  \citenamefont {Fetisov}, \citenamefont {Khomskii},\ and\ \citenamefont
  {Svozil}}]{Dolgov1987}%
  \BibitemOpen
  \bibfield  {author} {\bibinfo {author} {\bibfnamefont {O.~V.}\ \bibnamefont
  {Dolgov}}, \bibinfo {author} {\bibfnamefont {E.~P.}\ \bibnamefont {Fetisov}},
  \bibinfo {author} {\bibfnamefont {D.~I.}\ \bibnamefont {Khomskii}}, \ and\
  \bibinfo {author} {\bibfnamefont {K.}~\bibnamefont {Svozil}},\ }\bibfield
  {title} {\enquote {\bibinfo {title} {Model of interband pairing in mixed
  valence and heavy fermion systems},}\ }\href {\doibase 10.1007/BF01307308}
  {\bibfield  {journal} {\bibinfo  {journal} {Zeitschrift f\"ur Physik B
  Condensed Matter}\ }\textbf {\bibinfo {volume} {67}},\ \bibinfo {pages}
  {63--68} (\bibinfo {year} {1987})}\BibitemShut {NoStop}%
\bibitem [{\citenamefont {Agterberg}\ \emph {et~al.}(2017)\citenamefont
  {Agterberg}, \citenamefont {Shishidou}, \citenamefont {O'Halloran},
  \citenamefont {Brydon},\ and\ \citenamefont
  {Weinert}}]{PhysRevLett.119.267001}%
  \BibitemOpen
  \bibfield  {author} {\bibinfo {author} {\bibfnamefont {D.~F.}\ \bibnamefont
  {Agterberg}}, \bibinfo {author} {\bibfnamefont {T.}~\bibnamefont
  {Shishidou}}, \bibinfo {author} {\bibfnamefont {J.}~\bibnamefont
  {O'Halloran}}, \bibinfo {author} {\bibfnamefont {P.~M.~R.}\ \bibnamefont
  {Brydon}}, \ and\ \bibinfo {author} {\bibfnamefont {M.}~\bibnamefont
  {Weinert}},\ }\bibfield  {title} {\enquote {\bibinfo {title} {Resilient
  nodeless $d$-wave superconductivity in monolayer {FeSe}},}\ }\href {\doibase
  10.1103/PhysRevLett.119.267001} {\bibfield  {journal} {\bibinfo  {journal}
  {Phys. Rev. Lett.}\ }\textbf {\bibinfo {volume} {119}},\ \bibinfo {pages}
  {267001} (\bibinfo {year} {2017})}\BibitemShut {NoStop}%
\bibitem [{\citenamefont {Maier}\ \emph {et~al.}(2011)\citenamefont {Maier},
  \citenamefont {Graser}, \citenamefont {Hirschfeld},\ and\ \citenamefont
  {Scalapino}}]{PhysRevB.83.100515}%
  \BibitemOpen
  \bibfield  {author} {\bibinfo {author} {\bibfnamefont {T.~A.}\ \bibnamefont
  {Maier}}, \bibinfo {author} {\bibfnamefont {S.}~\bibnamefont {Graser}},
  \bibinfo {author} {\bibfnamefont {P.~J.}\ \bibnamefont {Hirschfeld}}, \ and\
  \bibinfo {author} {\bibfnamefont {D.~J.}\ \bibnamefont {Scalapino}},\
  }\bibfield  {title} {\enquote {\bibinfo {title} {$d$-wave pairing from spin
  fluctuations in the {${\mathrm{K}}_{x}$Fe${}_{2\ensuremath{-}y}$Se${}_{2}$}
  superconductors},}\ }\href {\doibase 10.1103/PhysRevB.83.100515} {\bibfield
  {journal} {\bibinfo  {journal} {Phys. Rev. B}\ }\textbf {\bibinfo {volume}
  {83}},\ \bibinfo {pages} {100515} (\bibinfo {year} {2011})}\BibitemShut
  {NoStop}%
\bibitem [{\citenamefont {Mazin}(2011)}]{PhysRevB.84.024529}%
  \BibitemOpen
  \bibfield  {author} {\bibinfo {author} {\bibfnamefont {I.~I.}\ \bibnamefont
  {Mazin}},\ }\bibfield  {title} {\enquote {\bibinfo {title} {Symmetry analysis
  of possible superconducting states in {K${}_{x}$Fe${}_{y}$Se${}_{2}$}
  superconductors},}\ }\href {\doibase 10.1103/PhysRevB.84.024529} {\bibfield
  {journal} {\bibinfo  {journal} {Phys. Rev. B}\ }\textbf {\bibinfo {volume}
  {84}},\ \bibinfo {pages} {024529} (\bibinfo {year} {2011})}\BibitemShut
  {NoStop}%
\bibitem [{\citenamefont {Kreisel}\ \emph {et~al.}(2013)\citenamefont
  {Kreisel}, \citenamefont {Wang}, \citenamefont {Maier}, \citenamefont
  {Hirschfeld},\ and\ \citenamefont {Scalapino}}]{PhysRevB.88.094522}%
  \BibitemOpen
  \bibfield  {author} {\bibinfo {author} {\bibfnamefont {A.}~\bibnamefont
  {Kreisel}}, \bibinfo {author} {\bibfnamefont {Y.}~\bibnamefont {Wang}},
  \bibinfo {author} {\bibfnamefont {T.~A.}\ \bibnamefont {Maier}}, \bibinfo
  {author} {\bibfnamefont {P.~J.}\ \bibnamefont {Hirschfeld}}, \ and\ \bibinfo
  {author} {\bibfnamefont {D.~J.}\ \bibnamefont {Scalapino}},\ }\bibfield
  {title} {\enquote {\bibinfo {title} {Spin fluctuations and superconductivity
  in {K${}_{x}$Fe${}_{2\ensuremath{-}y}$Se${}_{2}$}},}\ }\href {\doibase
  10.1103/PhysRevB.88.094522} {\bibfield  {journal} {\bibinfo  {journal} {Phys.
  Rev. B}\ }\textbf {\bibinfo {volume} {88}},\ \bibinfo {pages} {094522}
  (\bibinfo {year} {2013})}\BibitemShut {NoStop}%
\bibitem [{\citenamefont {Hu}(2013)}]{PhysRevX.3.031004}%
  \BibitemOpen
  \bibfield  {author} {\bibinfo {author} {\bibfnamefont {Jiangping}\
  \bibnamefont {Hu}},\ }\bibfield  {title} {\enquote {\bibinfo {title}
  {Iron-based superconductors as odd-parity superconductors},}\ }\href
  {\doibase 10.1103/PhysRevX.3.031004} {\bibfield  {journal} {\bibinfo
  {journal} {Phys. Rev. X}\ }\textbf {\bibinfo {volume} {3}},\ \bibinfo {pages}
  {031004} (\bibinfo {year} {2013})}\BibitemShut {NoStop}%
\bibitem [{\citenamefont {Chen}\ \emph {et~al.}(2020)\citenamefont {Chen},
  \citenamefont {Liu}, \citenamefont {Bao}, \citenamefont {Yan}, \citenamefont
  {Wang}, \citenamefont {Zhang},\ and\ \citenamefont
  {Feng}}]{PhysRevLett.124.097001}%
  \BibitemOpen
  \bibfield  {author} {\bibinfo {author} {\bibfnamefont {Chen}\ \bibnamefont
  {Chen}}, \bibinfo {author} {\bibfnamefont {Qin}\ \bibnamefont {Liu}},
  \bibinfo {author} {\bibfnamefont {Wei-Cheng}\ \bibnamefont {Bao}}, \bibinfo
  {author} {\bibfnamefont {Yajun}\ \bibnamefont {Yan}}, \bibinfo {author}
  {\bibfnamefont {Qiang-Hua}\ \bibnamefont {Wang}}, \bibinfo {author}
  {\bibfnamefont {Tong}\ \bibnamefont {Zhang}}, \ and\ \bibinfo {author}
  {\bibfnamefont {Donglai}\ \bibnamefont {Feng}},\ }\bibfield  {title}
  {\enquote {\bibinfo {title} {Observation of discrete conventional caroli--de
  gennes--matricon states in the vortex core of single-layer
  {$\mathrm{FeSe}/{\mathrm{SrTiO}}_{3}$}},}\ }\href {\doibase
  10.1103/PhysRevLett.124.097001} {\bibfield  {journal} {\bibinfo  {journal}
  {Phys. Rev. Lett.}\ }\textbf {\bibinfo {volume} {124}},\ \bibinfo {pages}
  {097001} (\bibinfo {year} {2020})}\BibitemShut {NoStop}%
\bibitem [{\citenamefont {Xiang}\ \emph {et~al.}(2025)\citenamefont {Xiang},
  \citenamefont {Wang},\ and\ \citenamefont {Wang}}]{Xiang2025Accidental}%
  \BibitemOpen
  \bibfield  {author} {\bibinfo {author} {\bibfnamefont {Ke}~\bibnamefont
  {Xiang}}, \bibinfo {author} {\bibfnamefont {Da}~\bibnamefont {Wang}}, \ and\
  \bibinfo {author} {\bibfnamefont {Qianghua}\ \bibnamefont {Wang}},\
  }\bibfield  {title} {\enquote {\bibinfo {title} {Accidental zero modes in a
  multiorbital superconductor with spin-orbital coupling},}\ }\href
  {https://pip.nju.edu.cn/CN/10.13725/j.cnki.pip.2025.05.003} {\bibfield
  {journal} {\bibinfo  {journal} {Progress in Physics}\ }\textbf {\bibinfo
  {volume} {45}},\ \bibinfo {pages} {250--259} (\bibinfo {year}
  {2025})}\BibitemShut {NoStop}%
\bibitem [{\citenamefont {Roig}\ \emph {et~al.}(2025)\citenamefont {Roig},
  \citenamefont {Islam}, \citenamefont {Oli}, \citenamefont {Zhang},
  \citenamefont {Brydon}, \citenamefont {Ramires}, \citenamefont {Yu},
  \citenamefont {Weinert}, \citenamefont {Li},\ and\ \citenamefont
  {Agterberg}}]{roig2025origin}%
  \BibitemOpen
  \bibfield  {author} {\bibinfo {author} {\bibfnamefont {Merc{\`e}}\
  \bibnamefont {Roig}}, \bibinfo {author} {\bibfnamefont {Kazi~Ranjibul}\
  \bibnamefont {Islam}}, \bibinfo {author} {\bibfnamefont {Basu~Dev}\
  \bibnamefont {Oli}}, \bibinfo {author} {\bibfnamefont {Huimin}\ \bibnamefont
  {Zhang}}, \bibinfo {author} {\bibfnamefont {PMR}\ \bibnamefont {Brydon}},
  \bibinfo {author} {\bibfnamefont {Aline}\ \bibnamefont {Ramires}}, \bibinfo
  {author} {\bibfnamefont {Yue}\ \bibnamefont {Yu}}, \bibinfo {author}
  {\bibfnamefont {Michael}\ \bibnamefont {Weinert}}, \bibinfo {author}
  {\bibfnamefont {Lian}\ \bibnamefont {Li}}, \ and\ \bibinfo {author}
  {\bibfnamefont {Daniel~F}\ \bibnamefont {Agterberg}},\ }\bibfield  {title}
  {\enquote {\bibinfo {title} {Origin of sublattice particle-hole asymmetry in
  monolayer {FeSe} superconductors},}\ }\href
  {https://arxiv.org/abs/2511.02226} {\bibfield  {journal} {\bibinfo  {journal}
  {arXiv:2511.02226}\ } (\bibinfo {year} {2025})}\BibitemShut {NoStop}%
\end{thebibliography}%

\pagebreak
\newpage
\clearpage
\onecolumngrid
\begin{center}
\textbf{\large Supplemental Material}
\end{center}
\setcounter{equation}{0}
\setcounter{figure}{0}
\setcounter{table}{0}
\setcounter{page}{1}
\makeatletter
\renewcommand{\theequation}{S\arabic{equation}}
\renewcommand{\thefigure}{S\arabic{figure}}
\renewcommand{\thetable}{S\arabic{table}}
\renewcommand{\bibnumfmt}[1]{[S#1]}
\renewcommand{\citenumfont}[1]{S#1}

\section{Details of the perturbation analysis}


This section is devoted to the perturbation analysis of Hamiltonian Eq.~\eqref{eq:BdG} under different pairing configurations. Firstly, we consider the case: $\Delta_{b1}=\Delta_{b2}=\Delta_b$ and $\Delta_{a1}=-\Delta_{a2}=\Delta_a$. When $\Delta_a=0$, the Hamiltonian is block-diagonal, with eigenvectors arranged column-wise as
\begin{equation}
    \begin{pmatrix}
        0 & -v_k & 0 & u_k \\
        -v_k & 0 & u_k & 0 \\
        u_k & 0 & v_k & 0 \\
        0 & u_k & 0 & v_k 
    \end{pmatrix}
\end{equation}
corresponding to eigenvalues $(-E^+_k, -E^-_k, E^-_k, E^+_k)$. The eigenvalues are defined as $E^\pm_k=\mathcal{E}_k\pm\eta_k=\sqrt{\epsilon_k^2+\Delta_b^2}\pm\eta_k$, where $\epsilon_k=\frac{\gamma_1+\gamma_2}{2}k^2-\mu$ and $\eta_k = \frac{\gamma_1-\gamma_2}{2}k^2$. In the vicinity of the Fermi surface, these four energy levels give rise to the coherence peaks in Fig.~\ref{fig1}(d). The BdG wavefunctions are defined as $u_k=\sqrt{\frac{1}{2}(1+\frac{\epsilon_k}{\mathcal{E}_k})}$ and $v_k=\sqrt{\frac{1}{2}(1-\frac{\epsilon_k}{\mathcal{E}_k})}$. Then we treat $\Delta_a$ as a perturbation. To first order, the eigenvectors become
\begin{equation}
    \begin{pmatrix}
        -\frac{\Delta_a}{2E^+_k}u_k & -v_k & \frac{\Delta_a}{2E^-_k}v_k & u_k \\
        -v_k & \frac{\Delta_a}{2E^-_k}u_k & u_k & -\frac{\Delta_a}{2E^+_k}v_k \\
        u_k & \frac{\Delta_a}{2E^-_k}v_k & v_k & \frac{\Delta_a}{2E^+_k}u_k \\
        -\frac{\Delta_a}{2E^+_k}v_k & u_k & -\frac{\Delta_a}{2E^-_k}u_k & v_k
    \end{pmatrix}.
\end{equation}
As shown in Fig.~\ref{fig2}(a), the two Fe sublattices are evenly mixed on the Fermi surfaces in the pristine normal state. We thus assume that $c^\dagger_{1k\sigma} = \frac{1}{\sqrt{2}}(c^\dagger_{Ak\sigma} + c^\dagger_{Bk\sigma}),\,\,
    c^\dagger_{2k\sigma} = \frac{1}{\sqrt{2}}(c^\dagger_{Ak\sigma} - c^\dagger_{Bk\sigma})$.
With this relation, we can compute the sublattice-projected DOS. Near the inner gap, the DOSs on sublattice $\mathrm{Fe_A}$ and $\mathrm{Fe_B}$ are 
\begin{equation}
    \begin{aligned}
        -E^-_k & \begin{cases}
            \rho_A(\omega) & \sim \left|\frac{\Delta_a}{2E^-_k}u_k-v_k\right|^2\delta(\omega+E^-_k),\\
            \rho_B(\omega) & \sim \left|\frac{\Delta_a}{2E^-_k}u_k+v_k\right|^2\delta(\omega+E^-_k);
        \end{cases}\\
        +E^-_k & \begin{cases}
            \rho_A(\omega) & \sim \left|\frac{\Delta_a}{2E^-_k}v_k+u_k\right|^2\delta(\omega-E^-_k),\\
            \rho_B(\omega) & \sim \left|\frac{\Delta_a}{2E^-_k}v_k-u_k\right|^2\delta(\omega-E^-_k).
        \end{cases}
    \end{aligned}
\end{equation}
If we assume $\Delta_a$ is negative, it is clear that the coherence peak at $-E^-_k$ on $\mathrm{Fe_A}$ is higher than that on $\mathrm{Fe_B}$ whereas the coherence peak at $+E^-_k$ on $\mathrm{Fe_A}$ is lower, which is exactly the feature of Fig.~\ref{fig1}(b). The situation for the outer gap, though not elaborated further here, likewise agrees with Fig.~\ref{fig1}(b).

Next is the second case: $\Delta_{b1}=\Delta_{b2}=\Delta_b$ and $\Delta_{a1}=\Delta_{a2}=\Delta_a$. Since the expressions for the perturbed eigenvectors in this case are rather cumbersome, we neglect terms proportional to $u_k^2 - v_k^2$, which are negligible near the Fermi surface. Within this approximation, the result to first order is 
\begin{equation}
    \begin{pmatrix}
        -\frac{2u_kv_k^2\Delta_a}{E^+_k-E^-_k} & -v_k & -\frac{2u_k^2v_k\Delta_a}{E^+_k-E^-_k} & u_k \\
        -v_k & \frac{2u_kv_k^2\Delta_a}{E^+_k-E^-_k} & u_k & \frac{2u_k^2v_k\Delta_a}{E^+_k-E^-_k} \\
        u_k & -\frac{2u_k^2v_k\Delta_a}{E^+_k-E^-_k} & v_k & \frac{2u_kv_k^2\Delta_a}{E^+_k-E^-_k} \\
        \frac{2u_k^2v_k\Delta_a}{E^+_k-E^-_k} & u_k & -\frac{2u_kv_k^2\Delta_a}{E^+_k-E^-_k} & v_k
    \end{pmatrix}.
\end{equation}
The sublattice-projected DOS is then
\begin{equation}
    \begin{aligned}
        -E^-_k & \begin{cases}
            \rho_A(\omega) & \sim \left|\frac{2u_kv_k^2\Delta_a}{E^+_k-E^-_k}-v_k\right|^2\delta(\omega+E^-_k),\\
            \rho_B(\omega) & \sim \left|\frac{2u_kv_k^2\Delta_a}{E^+_k-E^-_k}+v_k\right|^2\delta(\omega+E^-_k);
        \end{cases}\\
        +E^-_k & \begin{cases}
            \rho_A(\omega) & \sim \left|\frac{2u_k^2v_k\Delta_a}{E^+_k-E^-_k}-u_k\right|^2\delta(\omega-E^-_k),\\
            \rho_B(\omega) & \sim \left|\frac{2u_k^2v_k\Delta_a}{E^+_k-E^-_k}+u_k\right|^2\delta(\omega-E^-_k).
        \end{cases}
    \end{aligned}
\end{equation}
Unlike the previous case, where the sublattice contrast reversed between $-E^-_k$ and $+E^-_k$, the current result shows that sublattice $\mathrm{Fe_A}$ exhibits stronger coherence peaks at both energies, which is not the observed sublattice dichotomy.

Finally, we analyse the case: $\Delta_{b1}=-\Delta_{b2}=\Delta_b$ and no constraint on intraband pairing. Notably, in this case the BdG Hamiltonian Eq.~\eqref{eq:BdG} exhibits a symmetry
\begin{equation}
    \begin{pmatrix}
        0 & \sigma_z \\
        -\sigma_z & 0
    \end{pmatrix}H^*_{\mathrm{BdG}}\begin{pmatrix}
        0 & -\sigma_z \\
        \sigma_z & 0
    \end{pmatrix} = -H_{\mathrm{BdG}},
\end{equation}
where $\sigma_z$ is the Pauli matrix. If $H_{\mathrm{BdG}}$ has an eigenvector $(u_1,u_2,\dots,\dots)^T$ with energy $E$, then $(\dots,\dots,u_1^*,-u_2^*)^T$ is an eigenvector with energy $-E$. Owing to this symmetry, the DOS projected onto both sublattices is $(|u_1+u_2|^2+|u_1-u_2|^2)\delta(\omega-E)$, indicating no distinction between the two Fe sublattices.

In conclusion, only the case with interband pairing of the same sign and intraband pairing of opposite signs can give rise to the sublattice dichotomy.

\end{document}